\documentclass[showpacs,aps,prb,twocolumn,epsf,floatfix,%
superscriptaddress]{revtex4}
\usepackage{graphicx}
\usepackage{bm}
\usepackage{amsmath}
\begin{document}
\title{Theory of Excitonic States in CaB$_6$}
\author{Shuichi Murakami}
\email[Email: ]{murakami@appi.t.u-tokyo.ac.jp}
\affiliation{Department of Applied Physics, University of Tokyo,
Bunkyo-ku, Tokyo 113-8656, Japan}
\author{Ryuichi Shindou} 
\affiliation{Department of Applied Physics, University of Tokyo,
Bunkyo-ku, Tokyo 113-8656, Japan}
\author{Naoto Nagaosa}
\affiliation{Department of Applied Physics, University of Tokyo,
Bunkyo-ku, Tokyo 113-8656, Japan}
\affiliation{CERC, AIST Tsukuba Central 4, Tsukuba 305-8562, Japan}
\author{Andrei~S.~Mishchenko}
\affiliation{CERC, AIST Tsukuba Central 4, Tsukuba 305-8562, Japan}
\affiliation{RRC `Kurchatov Institute',123182, Moscow, Russia}
\date{\today}
\begin{abstract}
We study the excitonic states in CaB$_6$ in terms of the Ginzburg-Landau
theory. By minimizing the free energy and by comparing with experimental
results, we identify two possible ground states with exciton
condensation. They both break time-reversal and inversion symmetries. 
This leads to various magnetic and optical properties. 
As for magnetic properties, it is expected to be an antiferromagnet, and 
its spin structure is predicted. It will exhibit the magnetoelectric
effect, and observed novel ferromagnetism in doped samples and in 
thin-film and powder samples can arise from this effect. 
Interesting optical phenomena such as the nonreciprocal
 optical effect and the second harmonic generation are predicted. Their
measurement for CaB$_6$ will clarify 
whether exciton condensation occurs or not and 
which of the two states is realized.
\end{abstract}
\maketitle

\renewcommand{\labelenumi}{(\roman{enumi})}

\section{Introduction}
CaB$_6$ and its doped system Ca$_{1-x}$La$_{x}$B$_{6}$ have been the 
subject of intensive studies since the discovery of 
``high temperature ferromagnetism" \cite{young}.
The understanding of the parent compound CaB$_6$, 
which we focus 
in this paper, is 
essential to that of this novel ferromagnetism.
This ferromagnetism is peculiar in some respects: (i) the Curie
temperature ($\sim 600 {\rm K}$) is high
in spite of a small moment (0.07$\mu_{\rm B}$/La), (ii)
it occurs only in a narrow doping range ($x<0.01$), and (iii) there are 
no 
partially filled $d$- 
or $f$-bands.
Soon after this discovery, two scenarios are proposed;
one is based on a ferromagnetic phase of a dilute electron gas 
\cite{ceperley}, and
another is based on an excitonic state \cite{zra,bv} 
where the doped carriers are concluded to be fully spin-polarized 
\cite{vkr}. 
Earlier band structure calculations \cite{hy,mcpm} 
found the small overlap 
of the conduction and valence bands near the three X-points.
In the de Haas-van Alphen measurement,
the Fermi surfaces of both electrons and holes are observed \cite{hall},
which confirms the smallness of the gap.
Furthermore, the symmetry of the wave function at the bottom of 
conduction band (the top of the valence band) 
forbids a finite dipole 
matrix elements.
Therefore the dielectric constant is not enhanced even when the band gap is 
reduced \cite{hr},
and the excitonic instability is not suppressed; CaB$_{6}$
 can thus be regarded as
a promising candidate for the excitonic state.
A recent band calculation based on GW approximation \cite{tromp}, 
on the other hand, predicts a large band gap of $\sim$ 0.8 eV, which is too
large 
compared with the exciton binding energy of the order of $T_{c}\sim 600
{\rm K}$.
Measurements on photoemission \cite{ARPES}
and soft X-ray emission \cite{Xray}
observed a band gap of $\sim$ 1 eV, consistent with this GW
calculation; yet a surface effect in photoemission 
should be seriously considered before 
any conclusion is drawn.
Particularly, another GW calculation \cite{kino}
found the small overlap similar
to the LDA calculation.
Furthermore, there are several experimental evidences showing a symmetry 
breaking with magnetism below $T_{c}\sim 600$K .
X-ray diffraction and Raman scattering experiments \cite{udagawa}
shows an anomaly at
$T_{c}\sim 600$K, below which the symmetry is lowered from cubic to tetragonal.
A $\mu$SR measurement \cite{ohishi}
detected an internal magnetic field, which suggests 
an existence of a magnetic moment.

These findings contradict the large band gap because any instabilities
are unlikely if that is the case.
In addition, sensitivity to impurities and defects 
suggests a small band gap/overlap.
Thus, it seems likely that the exciton condensation occurs in this compound
due to a small band gap/overlap.
One can then attribute the high 
Curie temperature in Ca$_{1-x}$La$_{x}$B$_{6}$ to 
the excitonic instability in the parent compound, which is
determined by the exciton binding energy.
Nonetheless, even if we assume the exciton condensation, 
there remain many mysteries.
The peculiarity of 
ferromagnetism in Ca$_{1-x}$La$_{x}$B$_{6}$ reveals itself in 
the strong sample dependence.
In Ref.\ \onlinecite{zra}, 
the doped carriers are assumed to be essential to
ferromagnetism, but recent experiments could not find the correlation
between the transport properties and magnetism.
The smallness of the magnetic moment ($\sim 0.07\mu_{B}/$La)
is not fully accounted for, although
numerous attempts have been made \cite{zra,bg,b,vb,bbm}.
Moreover deficiency in Ca sites \cite{morikawa}
or doping of divalent elements 
such as Ba \cite{young} or Sr \cite{ott} also induces ferromagnetism.
Another hint is that the ferromagnetic moment is enhanced
in the thin-film sample \cite{terashima}, or near the surface 
from electron spin resonance (ESR) \cite{kunii}.
This fact suggests that some kind of symmetry lowering is related 
to the ferromagnetism.

Because of this controversial situation, it is 
worthwhile to give solid theoretical results which 
are independent of the details of microscopic 
models. In this paper we deal with this problem from 
the viewpoint of symmetry.
The only assumption we take as the basis 
of our analysis is that the parent compound CaB$_6$
is a triplet excitonic insulator with the order 
parameters made from the ${\rm X}'_3$- (conduction band)
and X$_3$- (valence band) wavefunctions 
at X-points. The analysis based on the 
Ginzburg-Landau expansion becomes exact by considering 
the symmetry properties of the order parameters.
All possible states are classified according to 
the irreducible representation of the magnetic point
group. We can then predict many nontrivial physical 
properties for each of the states,
which can be tested experimentally. Most important of 
these are that these states break
both the time reversal $(R)$ and space 
inversion $(I)$ symmetries, while their product
$RI$ remains intact. This implies that CaB$_{6}$ is an 
antiferromagnet (AF). This also leads to the 
magneto-electric (ME) effect; namely 
the ferromagnetic moment is induced when the 
electric field is applied or the electric 
polarization is induced when the magnetic field is 
applied. This offers a new mechanism of the
novel ferromagnetism in Ca$_{1-x}$La$_x$B$_6$, i.e.
ferromagnetism is induced near impurities or surfaces by the ME effect
We note that the present theory of ferromagnetism is different from
those in Ref.~\onlinecite{zra,bv}, though all these theories assume
exciton condensation.
The relation between our theory and the theories in 
Ref.~\onlinecite{zra,bv} is presented in detail in the last section.
We have published a short version \cite{msnm} of this paper,
but this paper contains new results such as a 
pattern of the magnetic moments, and optical
properties, as well as full and detailed 
description of the discussion. The plan of this 
paper is as follows. In Section II, the excitonic order parameters
are defined and the Ginzburg-Landau expansion 
is developed in III.  Physical properties of 
each state is described in Section IV, and V is
devoted to discussions and conclusions.

Here we remark a role of spin-orbit coupling. In subsequent discussions
we take into account the spin-orbit coupling. 
Its presence is crucial in our analysis by the GL theory.
Limit of zero spin-orbit coupling is discussed in Appendix 
\ref{appendix-a}, and it can be contrasted with the nonzero case.
In our numerical estimation for coefficients in GL free energy
(Fig.~\ref{2}), we 
neglected the spin-orbit coupling,
because it greatly simplifies the calculation; in interpreting its 
results, however, we should keep in 
mind that we are always treating the case of nonzero spin-orbit coupling.

\section{Excitonic Order Parameters}

Let $b_{\bm{k}\sigma}$ and 
$a_{\bm{k}\sigma}$ denote annihilation operators of electrons with spin 
$\sigma$ at the conduction and the valence bands respectively \cite{hr}.
When excitons are formed, excitonic order parameters 
$\langle b_{\bm{k}\alpha}^{\dagger}a_{\bm{k'}\beta}\rangle$
will have nonzero values. Excitonic instability occurs only in the 
vicinity of the X points \cite{hy,mcpm}, i.e.
$\bm{Q}_{x}=(\pi,0,0),\ \bm{Q}_{y}=(0,\pi,0),\ \bm{Q}_{z}=(0,0,\pi)$
in the cubic Brillouin zone, where the conduction and the valence bands 
approach each other with a small overlap/gap. 
We make the following assumptions;
\begin{enumerate}
 \item The order parameters are $\bm{k}$-independent near the X points.
 \item The order parameters connecting different X points are neglected
 \cite{bv}.
\end{enumerate}
Consequently, we keep only the order parameters 
$\eta(\bm{Q}_{i})_{\alpha\beta}=
\langle b_{\bm{Q}_{i}\alpha}^{\dagger}a_{\bm{Q}_{i}\beta}\rangle$
\cite{hrremark}.
Still we have $2\cdot 2\cdot 3=12$ complex order parameters involved
and the problem is quite complicated. The GL theory is so powerful that 
it helps us from this difficulty. The cubic (${\rm O}_{\rm h}$)
symmetry of the lattice considerably restricts the form of the GL free energy
$\Phi$, as is similar in the GL theory for 
unconventional superconductors \cite{vg,su}.
In this section and in the next section, we construct the GL free energy 
and discuss its mimima. The argument is rather technical,
because we will make full use of the point-group symmetry to deal with 
the twelve complex order parameters.
Readers who are not accustomed to detailed point-group discussions 
can jump to Section \ref{section-prediction}.

At each X point, the $\bm{k}$-group, i.e. the group which keeps $\bm{k}$ 
unchanged, is tetragonal (${\rm D}_{\rm 4h}$), 
and the conduction and the valence band states
belong to ${\rm X}^{\prime}_{3}$ and ${\rm X}_{3}$ representations,
respectively according to band calculation 
\cite{hy,mcpm}, when we take the origin  
as the center of a B$_6$ tetrahedron. In the absence of 
the spin-orbit coupling, 
the order parameter $\eta(\bm{Q}_{i})_{\alpha\beta}$ 
transform as a D$_{4\text{h}}$ irreducible representation
${\rm X}_{3} \times {\rm X}_{3}^{\prime}={\rm X}^{\prime}_{1}$.
From now on we shall take the spin-orbit coupling into account.
Then, the representation 
of $\eta$ will be altered. 
We restrict our analysis on the triplet channel for the excitons, because the 
exchange interaction usually favors the triplet state
 compared with the singlet state \cite{zra}.
Then, the spin-1 representation is multiplied to $\text{X}^{\prime}_{1}$, and 
the triplet order parameters follow the representations
${\rm X}_{4}^{\prime}+{\rm X}_{5}^{\prime}$ in the ${\rm D}_{{\rm 4h}}$
group. 
Let us now construct the order parameters explicitly.
We take the $\bm{Q}_{z}$ point as an
example; the triplet order parameters have three components, following 
the spin-1 representation.
By recalling that a spherical harmonics with $l=1$
is represented by $x\pm{\rm i}y,z$, we can recombine 
the triplet order parameters
as
\begin{eqnarray}
 &&\eta_{x}(\bm{Q}_{z})=-\frac{{\rm i}}{\sqrt{2}}
(\eta_{\uparrow\downarrow}(\bm{Q}_{z})+
\eta_{\downarrow\uparrow}(\bm{Q}_{z})), \label{etax}\\
 &&\eta_{y}(\bm{Q}_{z})=-\frac{{\rm 1}}{\sqrt{2}}
(\eta_{\uparrow\downarrow}(\bm{Q}_{z})-
\eta_{\downarrow\uparrow}(\bm{Q}_{z})), \label{etay}\\
 &&\eta_{z}(\bm{Q}_{z})=\frac{{\rm i}}{\sqrt{2}}
(\eta_{\downarrow\downarrow}(\bm{Q}_{z})-
\eta_{\uparrow\uparrow}(\bm{Q}_{z})), \label{etaz}
\end{eqnarray}
or in a short form
\begin{equation}
  \eta_{i}(\bm{Q}_{a})=-\frac{{\rm i}}{\sqrt{2}}
\sum_{\alpha,\beta}\eta_{\alpha\beta}(\bm{Q}_{a})\sigma^{i}_{\alpha\beta}.
\end{equation}
The reason why we introduced these linear combinations is because
they are convenient for subsequent symmetry discussions. 
Their phase factors are chosen so that the time-reversal $R$ operates as 
complex conjugation $R  \eta_{i}(\bm{Q}_{a})=
  \eta_{i}^{*}(\bm{Q}_{a})$.
By investigating their transformation property under D$_{4\text{h}}$,
which transforms $\bm{Q}_{z}$ onto itself, we find
that $\eta_{z}$ follows the $X'_{4}$
representation,
while $\eta_{x}$ and $\eta_{y}$ follow the $X'_{5}$ in ${\rm D}_{{\rm
4h}}$.
Here the spin-quantization axis in (\ref{etax})-(\ref{etaz}) 
is taken to be the $z$-axis.
Since we are taking into account the spin-orbit coupling, 
these up- and down-spins
should be interpreted as pseudospins \cite{su}.

When we consider the three X points, the cubic symmetry is restored.
We find that nine components of the 
order parameters are classified into irreducible representations of 
${\rm O}_{\rm h}$ as
${\rm \Gamma}_{\rm 15}+{\rm \Gamma}_{\rm 15}+
{\rm \Gamma}_{\rm 25}$, all of which are 3-dimensional.
Let us call basis functions as $\eta_{i}({\rm \Gamma}_{15},1)$,
$\eta_{i}({\rm \Gamma}_{15},2)$ and $\eta_{j}({\rm \Gamma}_{25})$,
where $i=x,y,z$ and $j=x(y^2 -z^2),y(z^2-x^2),z(x^2-y^2)$.
These basis functions transform like a vector $(x,y,z)$ for
$\Gamma_{15}$ and like $(x(y^2 -z^2),y(z^2-x^2),z(x^2-y^2))$ for 
$\Gamma_{25}$.
The basis functions are given as
\begin{eqnarray*}
&& \eta_{z}({\rm \Gamma}_{15},1)=
\eta_{z}(\bm{Q}_{z})=\frac{{\rm i}}{\sqrt{2}}
(\eta_{\downarrow\downarrow}(\bm{Q}_{z})-
\eta_{\uparrow\uparrow}(\bm{Q}_{z})), \\
&& \eta_{z}({\rm \Gamma}_{15},2)=
\frac{1}{\sqrt{2}}(\eta_{z}(\bm{Q}_{x})+
\eta_{z}(\bm{Q}_{y}))=
\nonumber \\
&&\makebox[1mm]{}
-\frac{1}{2}\{
{\rm i}(\eta_{\uparrow\downarrow}(\bm{Q}_{y})+
\eta_{\downarrow\uparrow}(\bm{Q}_{y}))
+(\eta_{\uparrow\downarrow}(\bm{Q}_{x})-
\eta_{\downarrow\uparrow}(\bm{Q}_{x}))
\}\ \ \ \ \ 
, \\
&& \eta_{z(x^2 -y^2)}({\rm \Gamma}_{25})=
\frac{1}{\sqrt{2}}(\eta_{z}(\bm{Q}_{x})-
\eta_{z}(\bm{Q}_{y}))=
\nonumber \\
&&\makebox[3mm]{}
\frac{1}{2}\{
{\rm i}(\eta_{\uparrow\downarrow}(\bm{Q}_{y})+
\eta_{\downarrow\uparrow}(\bm{Q}_{y}))
-(\eta_{\uparrow\downarrow}(\bm{Q}_{x})-
\eta_{\downarrow\uparrow}(\bm{Q}_{x}))
\}.
\end{eqnarray*} 
We defined these linear combinations in order 
to facilitate the construction of the GL free energy $\Phi$ in the 
next section.
Here, the spin-quantization axis in 
$\eta_{\alpha\beta}(\bm{Q}_{i})$ is taken as the 
$+i$-axis ($i=x,y,z$). 
Other components are obtained by cyclic permutations of $x,y,z$.
It is easily shown that $ I \bm{\eta} = - \bm{\eta} $, and 
$R \bm{\eta} = \bm{\eta}^* $, where we defined for brevity
\begin{eqnarray*}
&&\bm{\eta}(\Gamma_{15},1)=(\eta_{x}(\Gamma_{15},1),
\eta_{y}(\Gamma_{15},1), \eta_{z}(\Gamma_{15},1)),\\
&&\bm{\eta}(\Gamma_{15},2)=(\eta_{x}(\Gamma_{15},2),
\eta_{y}(\Gamma_{15},2), \eta_{z}(\Gamma_{15},2)),\\
&&\bm{\eta}(\Gamma_{25})=(\eta_{x(y^{2}-z^{2})}(\Gamma_{25}),
\eta_{y(z^{2}-x^{2})}(\Gamma_{25}), \\
&&\makebox[4cm]{}\eta_{z(x^{2}-y^{2})}(\Gamma_{25})).
\end{eqnarray*}

\section{Ginzburg-Landau Theory and Determination of the Ground State}
\subsection{Quadratic Order Terms}

Let us now write down the GL free energy in terms of these order
parameters. The GL free energy $\Phi$ should be invariant under the 
elements of ${\rm O}_{\rm h}$ and under the time-reversal $R$.
We make two remarks helpful in writing down $\Phi$.
First, only even-order terms in $\bm{\eta}$ are allowed
by the inversion symmetry. Second, owing to the time-reversal symmetry, 
the order of ${\rm Im}\bm{\eta}$ in each term should be even.
Thus, $\Phi$ is given up to quadratic order as
\begin{eqnarray}
 &&\Phi^{(2)}=A_{1}\sum_{i}({\rm Re}\ \eta_{i}(\Gamma_{15},1))^{2}
\nonumber \\
&&\makebox[.5mm]{}
+A_{2}\sum_{i}({\rm Re}\ \eta_{i}(\Gamma_{15},2))^{2}\ \ \ 
\nonumber \\
&&\makebox[.5mm]{}
+A_{3}\sum_{i}
{\rm Re}\ \eta_{i}(\Gamma_{15},1){\rm Re}\ \eta_{i}(\Gamma_{15},2)
\nonumber \\
&&\makebox[.5mm]{}
+A_{4}\sum_{j}({\rm Re}\ \eta_{j}(\Gamma_{25}))^{2}\nonumber \\
&&\makebox[.5mm]{}
+B_{1}\sum_{i}({\rm Im}\ \eta_{i}(\Gamma_{15},1))^{2}
+B_{2}\sum_{i}({\rm Im}\ \eta_{i}(\Gamma_{15},2))^{2}
 \nonumber \\
&&\makebox[.5mm]{}+
B_{3}\sum_{i}
{\rm Im}\ \eta_{i}(\Gamma_{15},1){\rm Im}\ \eta_{i}(\Gamma_{15},2)
\nonumber \\
&&\makebox[.5mm]{}
+B_{4}\sum_{j}({\rm Im}\ \eta_{j}(\Gamma_{25}))^{2}, \ \ \ \ \ \label{phi2}
\end{eqnarray}
where $A_{i},B_{i}$ are constants. As there exist no other symmetries
than considered above, these $A_{i}$ and $B_{i}$ are independent parameters.

Let us determine the ground state of the system from the 
GL free energy.
Minimizing $\Phi^{(2)}$, we find that one of the following states 
will be realized as the temperature is lowered;
\renewcommand{\labelenumi}{(\Roman{enumi})}
\begin{eqnarray*}
 \text{(A)}&& \text{Re}\ \bm{\eta}(\Gamma_{15},1) = c\ 
        \text{Re}\ \bm{\eta}(\Gamma_{15},2)\ne \bm{0},\\
 \text{(B)}&& 
{\rm Re}\ \bm{\eta}(\Gamma_{25})\neq \bm{0},\\ 
\text{(C)}&& {\rm Im}\ \bm{\eta}(\Gamma_{15},1) = c\ 
        {\rm Im}\ \bm{\eta}(\Gamma_{15},2)\ne \bm{0},\\
\text{(D)} &&{\rm Im}\ \bm{\eta}(\Gamma_{25})\neq \bm{0}, 
\end{eqnarray*}
where $c$ is a nonzero constant.
Therefore, condensations of excitons in $\Gamma_{15}$ and in
$\Gamma_{25}$ do not occur simultaneously.

In all cases (A)-(D), 
all directions of the vector $\bm{\eta}$ are degenerate, 
and this degeneracy will be lifted in the quartic order terms in $\Phi$, 
as we shall see afterwards. All these states are accompanied
by a lattice distortion. 
This distortion, however, 
is expected to be rather small, because it couples to the order 
parameter in the quadratic order, not linear order.
Indeed, the distortion detected by the X-ray scattering 
is tetragonal and is as small as $0.03 \%$ \cite{udagawa}.
This smallness of distortion is a common feature also in other hexaborides; 
for example, there observed no lattice distortion associated with 
low-temperature antiferromagnetism in GdB$_6$ \cite{gdb6}, 
and antiferroquadrupolar ordering in CeB$_6$ \cite{ceb6}.
Ferromagnetic EuB$_6$ \cite{eub6} should also have
a distortion away from a cubic lattice, though not observed experimentally.
It is because ferromagnetism cannot occur in a cubic 
lattice from general symmetry argument \cite{cubic}.

\subsection{Microscopic Calculation for Quadratic Order Terms}
To study which state is realized,
we shall calculate the coefficients in $\Phi^{(2)}$ by 
the Hartree-Fock approximation\cite{hr}, 
including coupling terms between the order parameters 
corresponding to the different X points. 
Because the spin-orbit coupling is weak in this compound, we
shall neglect it in the following microscopic calculation.

In order to derive the quadratic term $\Phi^{(2)}$ of the GL
theory for the triplet order parameters, we have only to concentrate on 
the following term\cite{hr}. 
\begin{eqnarray}
&&-\sum_{ij=x,y,z}(\langle a_{\bm{Q}_{i}\alpha}^{\dagger}
b_{\bm{Q}_{i}\beta}\rangle\delta_{\kappa,a}\delta_{\kappa',b}
+\langle b_{\bm{Q}_{i}\alpha}^{\dagger}
a_{\bm{Q}_{i}\beta}\rangle\delta_{\kappa,b}\delta_{\kappa',a}) \nonumber \\
&&\ \ \times V_{\bm{Q}_{j},\bm{Q}_{i}}^{\kappa\kappa'''
\kappa''\kappa'}
\delta_{\alpha,\eta}
\delta_{{\beta},{\gamma}}\times\nonumber \\
&&\ \ (\langle a_{\bm{Q}_{j}\gamma}^{\dagger}
b_{\bm{Q}_{j}\eta}\rangle\delta_{\kappa'',a}\delta_{\kappa''',b}
+\langle b_{\bm{Q}_{j}\gamma}^{\dagger}
a_{\bm{Q}_{j}\eta}\rangle\delta_{\kappa'',b}\delta_{\kappa''',a}),\label{1-1}
\end{eqnarray}
where 
\begin{eqnarray}
&&V_{\bm{Q}_{j},\bm{Q}_{i}}
^{\kappa\kappa'\kappa''\kappa'''}, \nonumber \\
&=&\int{\rm d}{\bm{x}}\int{\rm d}{\bm{x}'}
\phi_{\bm{Q}_{i}}^{\kappa}(\bm{x})\phi_{\bm{Q}_{j}}^{\kappa'}(\bm{x})
\frac{e^2}{\epsilon|\bm{x}-\bm{x}'|}
\phi_{\bm{Q}_{j}}^{\kappa''}(\bm{x}')\phi_{\bm{Q}_{i}}^{\kappa'''}(\bm{x}'),\nonumber 
\end{eqnarray}
and  $\epsilon$ is a dielectric 
constant, $\phi_{\bm{Q}_{i}}^{\kappa}(\bm{x})$ is the Bloch wavefunction of 
the $\kappa$-band
($\kappa = a,b$) at $\bm{k}=\bm{Q}_{i}$, taken as real.
We assume that screening of Coulomb interaction is weak. 
\begin{figure}[h]
\includegraphics[scale=0.35]{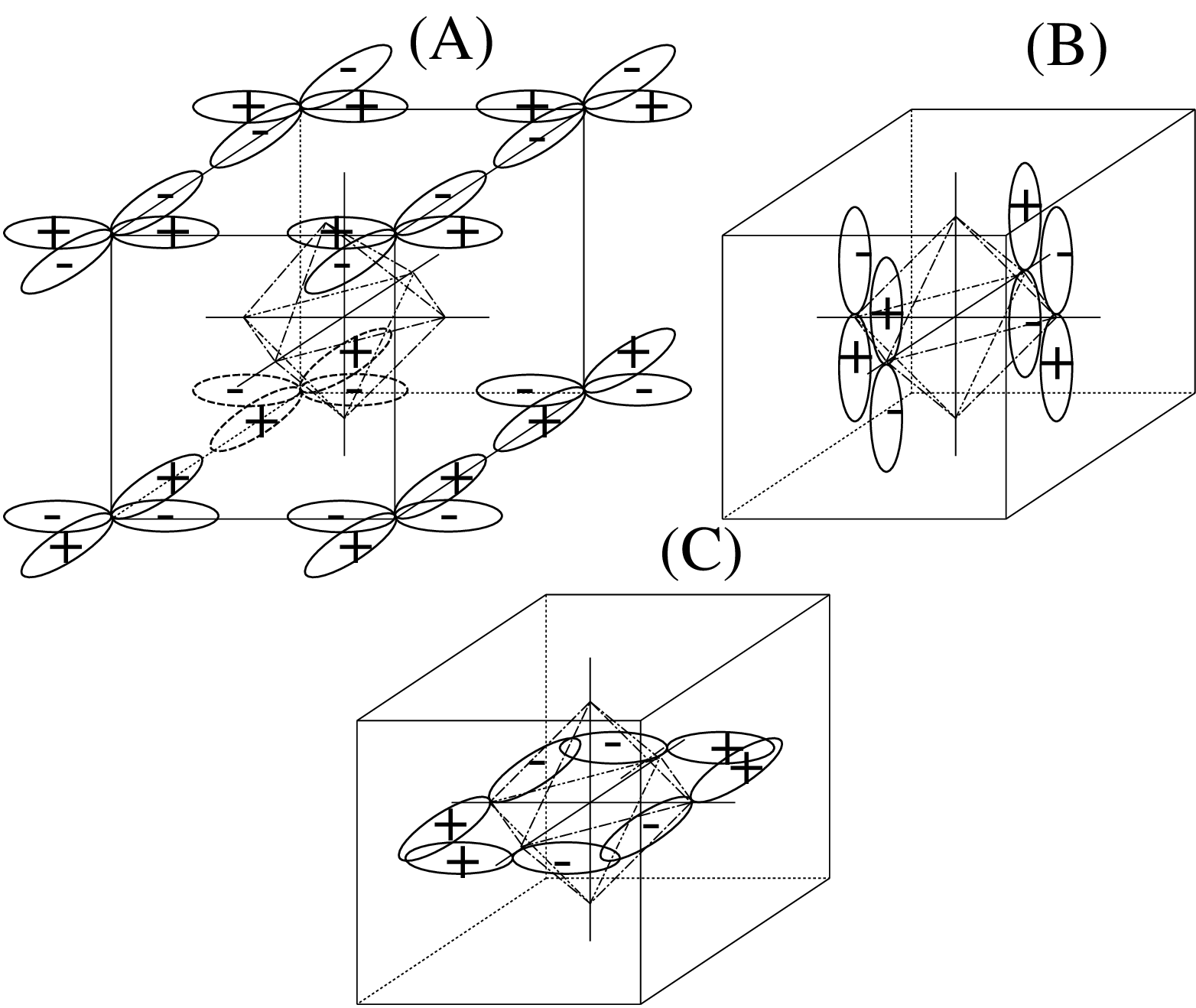}
\caption{(A) Wavefunction of $\rm{X}_{3}^{\prime}$ from the
d-orbit of the Ca-sites,
$\phi_{\bm{Q}_{z},{\rm X}_{3}^{\prime}}^{\rm{Ca:d-orbit}}$
(B) Wavefunction of ${\rm X}_{3}^{\prime}$ from the p-orbits of the 
B-sites,
$\phi_{\bm{Q}_{z},{\rm X}_{3}^{\prime}}^{\rm{B:p-orbit}}$
(C) Wavefunction of ${\rm X}_3$ from the p-orbits of the B-sites,
$\phi_{\bm{Q}_{z},{\rm X}_{3}}^{\rm{B:p-orbit}}$. All the wavefunctions
depicted in this figure have crystal momentum $\bm{Q}_{z}=(0,0,\pi)$ }
\label{orbit}
\end{figure}
Let us construct the wavefunctions involved in the exciton formation.
The wavefunction for the conduction band $\phi_{\bm{Q}_{i}}^{b}(\bm{x})$
at the X points mainly consists of the p-orbits of the boron ions
and the d-orbits of the calcium ions. 
These constituent wavefunctions, 
$\phi_{\bm{Q}_{i},{\rm X}_{3}'}^{\rm{B:p-orbit}}$ and 
$\phi_{\bm{Q}_{i},{\rm X}_{3}'}^{\rm{Ca:d-orbit}}$,
are depicted in Fig.\ \ref{orbit} (A) (B).
The wavefunction $\phi_{\bm{Q}_{i}}^{b}(x)$ is a
bonding orbital of these two: 
\begin{equation}
\phi_{\bm{Q}_{i}}^{b}(x)= 
c\phi_{\bm{Q}_{i},{\rm X}_{3}'}^{\rm{B:p-orbit}}
+(1-c)\phi_{\bm{Q}_{i},{\rm X}_{3}'}^{\rm{Ca:d-orbit}}.
\label{phia}
\end{equation}
On the other hand,
the wavefunction of the valence band at the X points is mainly
composed
of the p-orbits of the boron sites, as shown in Fig.\ \ref{orbit} (C).
As a simple estimation, 
we approximate the orbitals of the boron and the calcium ions as 
those of a hydrogen-like atom with charge $+2e$ and $+3e$,
respectively. We believe that using more realistic orbitals 
in the solid will not change our conclusions described below.
Owing to the time-reversal symmetry, 
the wavefunctions at each X-point can be taken as real.
Still, there
remains uncertainty in relative signs of the Bloch wavefunctions
among the X-points.  We define this as follows; 
\begin{eqnarray}
C_{3}^{[111]}\phi_{\bm{Q}_{z}}^{\kappa}(x)&=&\phi_{\bm{Q}_{x}}^{\kappa}
(x),\nonumber \\ 
(C_{3}^{[111]})^{2}\phi_{\bm{Q}_{z}}^{\kappa}(x)&=&
\phi_{\bm{Q}_{y}}^{\kappa}
(x),\nonumber
\end{eqnarray}
where $C_{3}^{[111]}$ is a three-fold 
rotation around the [111] direction.
This particular choice of relative signs of the wavefunctions at 
different X-points
does not affect physical consequences given below;
another choice leads to the same physical consequences.

Now we can evaluate the coefficient of eq.(\ref{phi2}) from eq.(\ref{1-1}).
Because there are some relation such as
 $V_{\bm{Q}_{x},\bm{Q}_{y}}
^{baba}=V_{\bm{Q}_{y},\bm{Q}_{x}}
^{baba}=V_{\bm{Q}_{y},\bm{Q}_{z}}
^{baba}=V_{\bm{Q}_{z},\bm{Q}_{x}}
^{baba}$, we have only to evaluate three quantities 
$\alpha=V_{\bm{Q}_{i},\bm{Q}_{i}}^{baba}$,
 $\beta=V_{\bm{Q}_{x},\bm{Q}_{y}}^{baba}$, $\gamma=
V_{\bm{Q}_{x},\bm{Q}_{y}}^{bbaa}$. 
In terms of these quantities, the coefficients in the 
GL free energy (\ref{phi2}) are expressed as 
\begin{equation}
\begin{array}{ll}
 A_{1}=C+\alpha, & A_{2}=C+\alpha+\beta-\gamma,
  \\
 A_{3}=2\sqrt{2}(\beta-\gamma), &A_{4}=C+\alpha-\beta+\gamma,
  \\
 B_{1}=C-\alpha, &B_{2}=C-\alpha-\beta-\gamma,
  \\
 B_{3}=-2\sqrt{2}(\beta+\gamma), 
 & B_{4}=C-\alpha+\beta+\gamma,
\end{array}
\end{equation}
where $C$ is a constant.
These satisfy relations
(\ref{no-spin-orbit})
in Appendix \ref{appendix-a},
which reflects the SU(2) symmetry in the spin space
in the limit of zero spin-orbit coupling.
To determine which type of excitons will condense, 
we diagonalize $\Phi^{(2)}$ as
\begin{eqnarray}
\Phi^{(2)}
&=&\lambda_{+}^{\rm{Re}}|{\rm{Re}}
\bm{\bm{\eta}}(\Gamma_{15},+)|^{2}
+\lambda_{+}^{\rm{Im}}|{\rm{Im}}
\bm{\bm{\eta}}(\Gamma_{15},+)|^{2}
\nonumber \\ 
&&+\lambda_{-}^{\rm{Re}}|{\rm{Re}}
\bm{\bm{\eta}}(\Gamma_{15},-)|^{2}
+\lambda_{-}^{\rm{Im}}|{\rm{Im}}
\bm{\bm{\eta}}(\Gamma_{15},-)|^{2}\nonumber \\
&&+\lambda_{\Gamma_{25}}^{\rm{Re}}
|{\rm{Re}}\bm{\bm{\eta}}(\Gamma_{25})|^{2}
+\lambda_{\Gamma_{25}}^{\rm{Im}}|{\rm{Im}}
\bm{\bm{\eta}}(\Gamma_{25})|^{2},
\label{1-2} 
\end{eqnarray}
where
\begin{eqnarray}
\lambda_{\pm}^{\rm{Re}}&=&C+\frac{1}{2}(2\alpha+\beta-\gamma
\pm 3(\beta-\gamma)),\nonumber \\
\lambda_{\pm}^{\rm{Im}}&=&C+\frac{1}{2}(-2\alpha-\beta-\gamma
\pm 3(-\beta-\gamma)),\nonumber \\
\lambda_{\Gamma_{25}}^{\rm{Re}}&=&C+
\alpha-\beta+\gamma=\lambda_{-}^{{\rm Re}},
\nonumber \\
\lambda_{\Gamma_{25}}^{\rm{Im}}&=&C
-\alpha+\beta+\gamma=\lambda_{-}^{{\rm Im}},
\label{lambdas}
\end{eqnarray}
and 
\begin{eqnarray}
&&\bm{\eta}(\Gamma_{15},+)=\frac{1}{\sqrt{3}}
(\bm{\eta}(\Gamma_{15},1)
+\sqrt{2}\bm{\eta}(\Gamma_{15},2)),\label{gamma+}\\
&&\bm{\eta}(\Gamma_{15},-)=\frac{1}{\sqrt{3}}
(\sqrt{2}\bm{\eta}(\Gamma_{15},1)
-\bm{\eta}(\Gamma_{15},2)).\label{gamma-}
\end{eqnarray}
The degeneracy in $\lambda_{\Gamma_{25}}^{\rm{Im}}
=\lambda_{-}^{\rm{Im}}$,\ 
$\lambda_{\Gamma_{25}}^{\rm{Re}}
=\lambda_{-}^{\rm{Re}}$  
is a consequence of (\ref{no-spin-orbit}) which is valid 
in the absence of the spin-orbit coupling; this degeneracy is lifted 
by the spin-orbit coupling.

\begin{figure}
\includegraphics[scale=0.3]{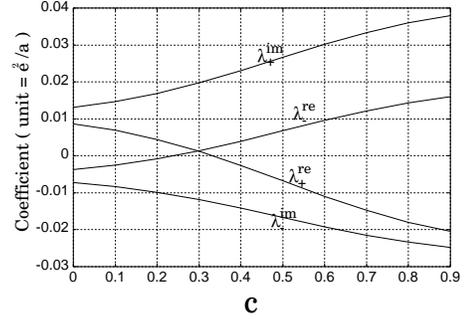}
\caption{Coefficients $\lambda_{\pm}^{\text{Re}}$,
$\lambda_{\pm}^{\text{Im}}$
of the GL free energy for real and imaginary
parts of the order parameters.
$a=$4.15$\text{\AA}$ is the lattice constant of CaB$_{6}$. 
The parameter $c$ controls the $\text{X}'_{3}$ wavefunctin of the 
conduction band. It is 
defined in (\ref{phia}) as a coefficient of mixing of the 
$p$-orbit of the boron and the 
$d$-orbit of the calcium.}
\label{2}
\end{figure}

Let us turn to numerical evaluation of $\alpha$, $\beta$, $\gamma$.
$\alpha$, $\beta$, $\gamma$ can be simplified as follows,
\begin{eqnarray}
\alpha&=&\sum_{\bm{G}{\ne}\bm{0}}\frac{1}{{\vert}\bm{G}\vert^{2}}
\left|
\int_{\rm{unit\hspace{1mm}cell}}d\bm{y}\tilde{\phi}_{\bm{Q}_{x}}^{a}(\bm{y})
\tilde{\phi}_{\bm{Q}_{x}}^{b}(\bm{y})e^{i\bm{G}\cdot\bm{y}}\right|^{2},
\nonumber \\
\beta&=&\sum_{\bm{G}}\frac{1}{{\vert}\bm{G}_{0}+\bm{G}\vert^{2}}
\int_{\rm{unit\hspace{1mm}cell}}d\bm{y}\tilde{\phi}_{\bm{Q}_{y}}^{b}(\bm{y})
\tilde{\phi}_{\bm{Q}_{x}}^{a}(\bm{y})e^{i(
\bm{G}_{0}+\bm{G})\cdot\bm{y}}\nonumber \\
&\times&
\int_{\rm{unit\hspace{1mm}cell}}d\bm{y}\tilde{\phi}_{\bm{Q}_{y}}^{a}(\bm{y})
\tilde{\phi}_{\bm{Q}_{x}}^{b}(\bm{y})e^{-i(
\bm{G}_{0}+\bm{G})\cdot\bm{y}},\nonumber \\
\gamma&=&\sum_{\bm{G}}\frac{1}{{\vert}
\bm{G}_{0}+\bm{G}\vert^{2}}
\int_{\rm{unit\hspace{1mm}cell}}d\bm{y}\tilde{\phi}_{\bm{Q}_{y}}^{a}(\bm{y})
\tilde{\phi}_{\bm{Q}_{x}}^{a}(\bm{y})e^{i(
\bm{G}_{0}+\bm{G})\cdot\bm{y}}\nonumber \\
&\times&
\int_{\rm{unit\hspace{1mm}cell}}d\bm{y}\tilde{\phi}_{\bm{Q}_{y}}^{b}(\bm{y})
\tilde{\phi}_{\bm{Q}_{x}}^{b}(\bm{y})e^{-i(\bm{G}_{0}+\bm{G})\cdot\bm{y}},
\nonumber 
\end{eqnarray}
where $\bm{G}_{0}={}^{\rm t}(\pi,\ \pi,\ 0)$, and 
the summation over $\bm{G}$ represents the one over 
reciprocal vectors.
The function $\tilde{\phi}_{\bm{Q}_{i}}^{c}(\bm{x})$ is
a localized atomic wavefunction, and we approximate it as
\begin{equation}
 \tilde{\phi}_{\bm{Q}_{i}}^{c}(\bm{x})\sim\left\{
\begin{array}{ll}
 \sqrt{N}\phi_{\bm{Q}_{i}}^{c}(\bm{x})& \mbox{: inside the unit cell,}\\
0 & \mbox{: outside the unit cell.}\\
\end{array}
\right.
\end{equation}
Hence, it satisfies
$\phi_{\bm{Q}_{i}}^{c}(\bm{x})=\frac{1}{\sqrt{N}}
\sum_{\bm{T}}\tilde{\phi}_{\bm{Q}_{i}}^{c}(\bm{x-T})e^{i\bm{T}
\cdot\bm{Q}_{i}}$, where $\bm{T}$ denotes a translational vector
of the lattice.
We have omitted the factor $e^{2}/\epsilon$ in the expressions of
$\alpha,\ \beta,\ \gamma$, since it is irrelevant for the present 
discussion.
Shown in Fig.\ref{2} are the coefficients of eq.(\ref{1-2}) 
as a function of $c$ defined in (\ref{phia}). 
The constant part $C$ in (\ref{lambdas}) common among these 
coefficients is omitted in this calculation, since
it is unnecessary for comparison of $\lambda^{{\rm Im}}$  and
$\lambda^{{\rm Re}}$.
As is shown in Fig.\ref{2}, the coefficients involving 
the imaginary parts are smaller than those for the real parts.
Therefore, the states (C) and (D) would be favorable than 
(A) or (B). 
In the notation in Ref.\ \onlinecite{hr},
the type-II and type-III is more favorable
than the type-IV and type-I.
The states (C)(D) breaks the time-reversal symmetry. Thus 
they are magnetic states, which seems to fit an appearance of
ferromagnetism in several conditions like powder 
or thin film in CaB$_6$.
On the other hand, the states (A)(B) preserve the time-reversal 
symmetry and thus nonmagnetic. From symmetry consideration,
this implies that neither the ME nor the 
piezomagnetic (PM) effect will be observed 
\cite{birss}.
Roughly speaking these states are far from showing ferromagnetism.
Thus, hereafter we shall concentrate on (C) and (D) \cite{hr-comment}.

Figure \ref{2} shows $\lambda_{+}^{{\rm Im}}>\lambda_{-}^{{\rm Im}}=
\lambda_{\Gamma_{25}}^{{\rm Im}}$, meaning that 
the condensation of excitons 
in $\Gamma_{15}$ and those in $\Gamma_{25}$ seem to occur simultaneously.
This is an artifact of approximation of zero spin-orbit coupling. 
We note that we are working with the presence of spin-orbit coupling;
these two condensations do not occur simultaneously, 
and either (C) or (D) is selected.

\subsection{Quartic Order Terms}

Let us consider quartic order terms $\Phi^{(4)}$ in the GL free energy, 
which lifts the degeneracy in the direction of $\bm{\eta}$.
The terms in $\Phi^{(4)}$ containing only the imaginary parts of the 
order parameters are written as
\begin{eqnarray*}
 &&\Phi^{(4)}=\\
&&\makebox[2mm]{}
 D_{1}\left\{\sum_{i}({\rm Im}\ \eta_{i}(\Gamma_{15},1))^{2}\right\}^{2}
+E_{1}\sum_{i}({\rm Im}\ \eta_{i}(\Gamma_{15},1))^{4}\nonumber \\
&&\makebox[2mm]{}
+D_{2}\left\{\sum_{i}({\rm Im}\ \eta_{i}(\Gamma_{15},2))^{2}\right\}^{2}
+E_{2}\sum_{i}({\rm Im}\ \eta_{i}(\Gamma_{15},2))^{4}\nonumber \\
&&\makebox[2mm]{}
+D_{3}\left\{\sum_{i}({\rm Im}\ \eta_{i}(\Gamma_{25}))^{2}\right\}^{2}
+E_{3}\sum_{i}({\rm Im}\ \eta_{i}(\Gamma_{25}))^{4}\nonumber \\
&&\makebox[2mm]{}
+F_{1}\left\{\sum_{i}({\rm Im}\ \eta_{i}(\Gamma_{15},1))^{2}\right\}
\left\{\sum_{i}({\rm Im}\ \eta_{i}(\Gamma_{15},2))^{2}\right\} \nonumber \\
&&\makebox[2mm]{}
+F_{2}\left\{
3(
{\rm Im}\ \eta_{x}(\Gamma_{15},1)^{2}-
{\rm Im}\ \eta_{y}(\Gamma_{15},1)^{2})\right.\nonumber \\
&&\makebox[6mm]{}
\times ({\rm Im}\ \eta_{x}(\Gamma_{15},2)^{2}-
{\rm Im}\ \eta_{y}(\Gamma_{15},2)^{2})\nonumber \\
&&\makebox[2mm]{}
+
(
2{\rm Im}\ \eta_{z}(\Gamma_{15},1)^{2}-
{\rm Im}\ \eta_{x}(\Gamma_{15},1)^{2}-
{\rm Im}\ \eta_{y}(\Gamma_{15},1)^{2})\nonumber \\
&&\makebox[6mm]{}
\left.
\times(
2{\rm Im}\ \eta_{z}(\Gamma_{15},2)^{2}-
{\rm Im}\ \eta_{x}(\Gamma_{15},2)^{2}-
{\rm Im}\ \eta_{y}(\Gamma_{15},2)^{2})
\right\}
\nonumber \\
&&\makebox[2mm]{}
+F_{3}\nonumber\\
&&\makebox[4mm]{}
\times\left(
{\rm Im}\ \eta_{x}(\Gamma_{15},1)
{\rm Im}\ \eta_{y}(\Gamma_{15},1)
{\rm Im}\ \eta_{x}(\Gamma_{15},2)
{\rm Im}\ \eta_{y}(\Gamma_{15},2)\right.\nonumber \\
&&\makebox[2mm]{}
+{\rm Im}\ \eta_{y}(\Gamma_{15},1)
{\rm Im}\ \eta_{z}(\Gamma_{15},1)
{\rm Im}\ \eta_{y}(\Gamma_{15},2)
{\rm Im}\ \eta_{z}(\Gamma_{15},2)\nonumber \\
&&\makebox[2mm]{}
+\left.
{\rm Im}\ \eta_{z}(\Gamma_{15},1)
{\rm Im}\ \eta_{x}(\Gamma_{15},1)
{\rm Im}\ \eta_{z}(\Gamma_{15},2)
{\rm Im}\ \eta_{x}(\Gamma_{15},2)
\right)\nonumber \\
&&\makebox[2mm]{}
+\sum_{i}
{\rm Im}\ \eta_{i}(\Gamma_{15},1)
{\rm Im}\ \eta_{i}(\Gamma_{15},2)\nonumber \\
&&\makebox[6mm]{}
\times\left\{
G_{1}
{\rm Im}\ \eta_{i}(\Gamma_{15},1)^{2}
+G_{2}
\sum_{i'}{\rm Im}\ \eta_{i'}(\Gamma_{15},1)^{2}
\right.\nonumber \\
&&\makebox[6mm]{}
\left.
+G_{3}
{\rm Im}\ \eta_{i}(\Gamma_{15},2)^{2}
+G_{4}
\sum_{i'}{\rm Im}\ \eta_{i'}(\Gamma_{15},2)^{2}
\right\}.
\end{eqnarray*}
Here coupling terms between $\Gamma_{15}$ and $\Gamma_{25}$ are
not written, because we are dealing only with condensation of 
one type of excitons.
It is noted here that the coupling of the order parameter $\eta$ and the 
lattice distortion $u$ such as $u^2 \eta^2$ also produces the quartic terms in
$\eta$ after integrating over $u$, and hence lifts the degeneracy.
This effect is included in $\Phi^{(4)}$.
By minimizing $\Phi^{(2)}+\Phi^{(4)}$,
we find four possibilities;
\begin{eqnarray*}
\text{(C-1)}&& {\rm Im}\ \bm{\eta}(\Gamma_{15},1) = c_1 
        {\rm Im}\ \bm{\eta}(\Gamma_{15},2) = (0,0,c_2) ,\\
\text{(C-2)}&& {\rm Im}\ \bm{\eta}(\Gamma_{15},1) = c_1
        {\rm Im}\ \bm{\eta}(\Gamma_{15},2)= (c_2,c_2,c_2) ,\\
\text{(D-1)}&&{\rm Im}\ \bm{\eta}(\Gamma_{25}) = (0,0,c_{2}),\\
\text{(D-2)} &&{\rm Im}\ \bm{\eta}(\Gamma_{25})=
(c_{2},c_{2},c_{2}),
\end{eqnarray*}
where $c$'s are constants.
There are three equivalent directions 
[001] for (C-1) and (D-1) and four directions [111] for (C-2) and (D-2);
we take ony one of them without loss of generality.
The  direction of the lattice distortion is 
tetragonal in (C-1) and (D-1) and is trigonal 
in (C-2) and (D-2).
We remark here on the recent results of 
X-ray scattering and Raman scattering \cite{udagawa},
which strongly support our theory. They show 
a tetragonal distortion below 600K, which indicates that (C-1) or (D-1)
is realized in CaB$_6$. Furthermore, 
inversion-symmetry breaking does not manifest itself in the Raman 
spectrum \cite{udagawa2}, which is consistent with (C) or (D), but not
with (A) or (B). The reason is the following.
In all cases (A)-(D) the inversion symmetry is broken. 
In (A) and (B) this broken inversion symmetry will appear in the 
Raman spectrum. The cases (C)(D), on the other hand, are
invariant under an $RI$ symmetry, as we shall see later. 
Since the phonon spectrum is time-reversal invariant, it is 
automatically invariant under $RI\cdot R =I$, i.e. inversion
symmetric. Thus, although (C)(D)
breaks inversion symmetry, it does not appear in the Raman spectrum.
Therefore, the phonon spectrum, which is even in time-reversal, 
automatically become even in inversion. Thus, although (C)(D)
breaks inversion symmetry, it does not appear in the Raman spectrum.

To summarize, (C-1) and (D-1) are the only possibilities 
totally consistent with the X-ray and Raman scattering\cite{udagawa}.
In the states (C-1) and (D-1), the condensed excitons in each case are 
given as 
\begin{eqnarray}
\mbox{(C-1)} &:& 
\bm{\eta}(\bm{Q}_{x})
=\bm{\eta}(\bm{Q}_{y})
=(0,0,{\rm i}c_{4}),\nonumber \\
&&
\bm{\eta}(\bm{Q}_{z})
=(0,0,{\rm i}c_{5}), \label{C1eta}\\
\mbox{(D-1)}&:& 
\bm{\eta}(\bm{Q}_{x})
=-\bm{\eta}(\bm{Q}_{y})
=(0,0,{\rm i}c_{6}),\nonumber \\
&&
\bm{\eta}(\bm{Q}_{z})
=0,\label{D1eta}
\end{eqnarray}
where $c_{i}$ are real. To be explicit, we can rewrite them as
\begin{eqnarray}
\mbox{(C-1)} &:& 
\langle b^{\dagger}_{\bm{Q}_{z}\downarrow}a_{\bm{Q}_{z}\downarrow}\rangle
=-\langle 
b^{\dagger}_{\bm{Q}_{z}\uparrow}a_{\bm{Q}_{z}\uparrow}\rangle =
\frac{c_{5}}{\sqrt{2}}, \\
&&
-\langle b^{\dagger}_{\bm{Q}_{y}\uparrow}a_{\bm{Q}_{y}\downarrow}\rangle
=-\langle 
b^{\dagger}_{\bm{Q}_{y}\downarrow}a_{\bm{Q}_{y}\uparrow}\rangle \nonumber 
\\
&&
={\rm i}\langle 
b^{\dagger}_{\bm{Q}_{x}\uparrow}a_{\bm{Q}_{x}\downarrow}\rangle =
-{\rm i}\langle 
b^{\dagger}_{\bm{Q}_{x}\downarrow}a_{\bm{Q}_{x}\uparrow}\rangle =
\frac{c_{4}}{\sqrt{2}},\\
\mbox{(D-1)}&:& 
\langle b^{\dagger}_{\bm{Q}_{y}\uparrow}a_{\bm{Q}_{y}\downarrow}\rangle
=\langle 
b^{\dagger}_{\bm{Q}_{y}\downarrow}a_{\bm{Q}_{y}\uparrow}\rangle \nonumber \\
&&
={\rm i}\langle 
b^{\dagger}_{\bm{Q}_{x}\uparrow}a_{\bm{Q}_{x}\downarrow}\rangle =
-{\rm i}\langle 
b^{\dagger}_{\bm{Q}_{x}\downarrow}a_{\bm{Q}_{x}\uparrow}\rangle =
\frac{c_{6}}{\sqrt{2}},
\end{eqnarray}
These expressions are too complicated to extract
information for physical properties; instead our approach is based on 
symmetry, namely,  magnetic point groups.
Magnetic point groups $G$ for these states are easily obtained, 
if we consider which symmetry operations including $R$
keeps the values of the order parameters unchanged. The results are
\begin{eqnarray*}
&&\mbox{(C-1):} \ \ G=   4/{\rm m}'{\rm mm}={\rm C}_{{\rm 4v}}\times\{E,RI\} \\
&&
\makebox[1.4cm]{}
=\{ E,C_{4z}^{\pm},C_{2z},\sigma_{x},
\sigma_{y},\sigma_{da},\sigma_{db}\}\times \{E,RI\},\\
&&\mbox{(D-1):} \ \  G= 
4'/{\rm m}'{\rm m}'{\rm m}={\rm D}_{{\rm 2d}}\times\{E,RI\} \\
&&\makebox[1.4cm]{}
=\{ E,S_{4z}^{\pm},C_{2z},C_{2a},
C_{2b},\sigma_{x},\sigma_{y}\}\times \{E,RI\}.
\end{eqnarray*}
Here symmetry operations are defined as follows;
$E$: identity; 
$C_{2j}$ ($j=x,y,z$); two-fold rotation around the $j$-axis; 
$C_{2a}$, $C_{2b}$: two-fold rotation around the $[110]$ and
 $[1\bar{1}0]$ axes, respectively;
$C_{4z}^{\pm}$: $\pm\pi/2$ rotation around the $z$-axis;
$\sigma_{j}$ ($j=x,y,z$): reflection with respect to the plane normal 
to the $j$-axis; 
$S_{4z}^{\pm}=C_{4z}^{\pm}I$;  
$\sigma_{da}$, $\sigma_{db}$:  
reflection with respect to the $(110)$ plane and
the $(1\bar{1}0)$ plane, respectively.
These will be used in the next section to make various predictions 
of the present material.
In (C-1) and (D-1), when we fix the axis of tetragonal distortion, 
there are two types of degenerate AF
domains, related with each other by the time-reversal.
Since there are three choices of axes for the tetragonal axis, 
total degeneracy is six in (C-1)(D-1), 
which is equal to an order of the quotient group 
$({\rm O}_{\rm h}\times\{ E,R \})/G$.
We can draw some analogies with anisotropic superconductivity (SC).
The order parameters {\boldmath$\eta$} of triplet excitons correspond 
to the d-vector in triplet SC. It is nevertheless misleading to 
look for SC counterparts of our phases (C-1) (D-1), because our order 
parameters are confined in the neighborhood of the three X points.
They are triplet and even functions in ${\bf k}$, which never occurs 
in the SC.

\section{Prediction of Physical Properties}
\label{section-prediction}
\subsection{Magnetic Properties}

A crucial observation for prediction of physical properties 
of these states is that they do not break the $RI$ symmetry,
while both $R$ and $I$ are broken \cite{zra}. 
There are a number of compounds, such as Cr$_2$O$_3$, known to
possess this feature, leading to several magnetic properties 
listed below.
First, the $RI$ symmetry prohibits a presence of any
uniform magnetization.
Dzyaloshinskii used this symmetry to explain
why  weak ferromagnetism is present in $\alpha$-Fe$_2$O$_3$ while not in
Cr$_2$O$_3$ \cite{dzy1}.
Thus the states (C-1) and (D-1) are antiferromagnetic.
This agrees with the result of the $\mu$SR measurement \cite{ohishi}
with a moment of $0.0039\mu_{{\rm B}}/{\rm mol}$.
Note that the magnetic unit cell is identical
with the original unit cell. Thus,
no extra Bragg spots appear below the AF phase transition.
Second, the $RI$ symmetry also prohibits 
the piezomagnetic (PM) effect. \cite{PMnote} 
A uniform stress cannot break the $RI$ 
symmetry which hinders ferromagnetism.
Nevertheless, a {\it gradient} of stress can break this 
symmetry and will induce ferromagnetism.

Third, the $RI$-invariance results in the linear ME effect,
as is first
observed in Cr$_2$O$_3$ \cite{dzy2,odell}. 
The $RI$-invariance allows the free energy to have a term like 
$G_{ij} H_{i}E_{j}$, causing the 
magnetization $\bm{M}$ proportional to $\bm{E}$, and 
the polarization $\bm{P}$ proportional to $\bm{H}$:
\begin{equation}
 \bm{M}=\bm{G}\bm{E},\ \ 
 \bm{P}={}^{t}\bm{G}\bm{H}.
\end{equation}
Roughly speaking, this occurs because an external electric field $\bf E$
breaks this $RI$ symmetry and enables ferromagnetism \cite{dzy2}.
In the GL language, this can be stated as follows. 
An electric field $\bf E$ belongs to the ${\rm \Gamma}_{15}$ representation, 
and couples linearly with the order parameters in (C-1) as
\begin{equation}
 \delta\Phi =
-C_{1}\bm{E}\cdot{\rm Re}\ \bm{\eta}(\Gamma_{15},1)
-C_{2}\bm{E}\cdot {\rm Re}\ \bm{\eta}(\Gamma_{15},2),
\end{equation}
in the lowest order.
Here the imaginary parts of the order parameters are absent
due to invariance of $\delta\Phi$ under time-reversal.
Thus, in the presence of $\bm{E}$, both the real and imaginary 
parts of the order parameters acquire nonvanishing values; this 
breaks the $RI$ symmetry, resulting 
in a ferromagnetic moment.
As for the (D-1), similar effect can be found
in
\begin{eqnarray}
&& \delta\Phi \nonumber \\
&&\ =-C_{1}
\sum_{i=1}^{3}
({\rm Im} \eta_{i}(\Gamma_{25})^{2}(3{\rm Re}\eta_{i}(\Gamma_{25})E_{i}
-{\rm Re}\bm{\eta}(\Gamma_{25})\cdot\bm{E}))\nonumber \\
&&\  -C_2
({\rm Im} \eta_{1}(\Gamma_{25}){\rm Im} \eta_{2}(\Gamma_{25})
({\rm Re}\bm{\eta}(\Gamma_{25})\times\bm{E})_{3}
\nonumber \\
&&\ \  +{\rm Im} \eta_{2}(\Gamma_{25}){\rm Im} \eta_{3}(\Gamma_{25})
({\rm Re}\bm{\eta}(\Gamma_{25})\times\bm{E})_{1}
\nonumber \\
&&\ \  +{\rm Im} \eta_{3}(\Gamma_{25}){\rm Im} \eta_{1}(\Gamma_{25})
({\rm Re}\bm{\eta}(\Gamma_{25})\times\bm{E})_{2}),
\end{eqnarray}
where $\bm{\eta}(\Gamma_{25})=(\eta_{1},\eta_{2},\eta_{3})$. 
Below the exciton condensation temperature, ${\rm Im} 
\bm{\eta}
(\Gamma_{25})$
has a nonvanishing value, which in turn brings about
a linear coupling between ${\rm Re}\bm{\eta}
(\Gamma_{25})$ and $\bm{E}$.

\begin{figure}
 \includegraphics[scale=0.45]{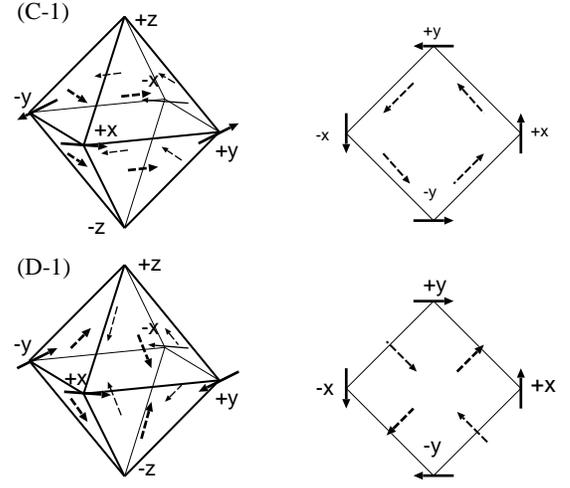}
 \caption{Schematic picture of spin structure for 
 (C-1) and (D-1). The right panels in (C-1) and (D-1) show the 
 top views, i.e. the views from the [001]-direction.}
 \label{spin}
 \end{figure}

These properties described above are deduced solely from the $RI$-invariance, and 
are common in (C-1) and (D-1). Meanwhile, detailed magnetic properties 
vary among them.
Let us first investigate a AF magnetic structure within the 
unit cell. We can calculate 
possible magnetic structure consistent with each magnetic point group. 
We cannot determine 
the quantitative distribution of magnetic moments. Their distribution can
be novel, since in the isostructural compound CeB$_6$, polarized 
neutron scattering 
shows that magnetic moments are at three locations: at the Ce sites, at the 
centers of the triangular plaquettes of B$_6$ octahedra, and 
at the midst of the 
B-B links connecting neighboring octahedra. \cite{saitoh}
Anyway, to look at the difference intuitively, we displayed 
in Fig. \ref{spin} 
directions of magnetic moments both at the boron sites and 
at the centers of the triangular plaquettes on B$_6$ octahedra 
with the magnitudes arbitrary chosen.
Let us call the six moments at each boron site as $\bm{S}_{\pm x}$,
$\bm{S}_{\pm y}$, $\bm{S}_{\pm z}$;
the obtained magnetic structure is 
\begin{eqnarray}
 \mbox{(C-1)}&:&\bm{S}_{\pm x}=\pm(0,a_{1},0),\ \bm{S}_{\pm y}=
\mp(a_{1},0,0),
\nonumber \\
&&
\bm{S}_{\pm z}=\bm{0},\label{spinC1}\\
 \mbox{(D-1)}&:&\bm{S}_{\pm x}=\pm(0,a_{1},0),\ 
\bm{S}_{\pm y}=\pm(a_{1},0,0),
\nonumber \\
&&
\bm{S}_{\pm z}=\bm{0},\label{spinD1}
\end{eqnarray}
where $a_{1}$ 
is a constant. 
On the other hand, Let $\bm{S}_{\alpha\beta\gamma}$ 
($\alpha, \beta, \gamma=\pm$) denote a moment on a triangular 
plaquette with vertices $\bm{S}_{\alpha x}$,
$\bm{S}_{\beta y}$, and 
$\bm{S}_{\gamma z}$. Then they are given by\begin{eqnarray*}
&& \mbox{(C-1)}:\bm{S}_{\pm\pm\pm}=\pm a_{3}(-1,1,0),\ \bm{S}_{\mp\pm\pm}=
\pm a_{3}(-1,-1,0),
\nonumber \\
&& \makebox[1cm]{}
\bm{S}_{\pm\mp\pm}=\pm a_{3}(1,1,0),\ \bm{S}_{\pm\pm\mp}=
\pm a_{3}(-1,1,0),\\
&& \mbox{(D-1)}:
\bm{S}_{\pm\pm\pm}=\pm (a_{3},a_{3},-a_{2}),\ \bm{S}_{\mp\pm\pm}=
\pm (a_{3},-a_{3},a_{2}),
\nonumber \\
&& \makebox[1cm]{}
\bm{S}_{\pm\mp\pm}=\pm (-a_{3},a_{3},a_{2}),\ \bm{S}_{\pm\pm\mp}=
\pm (a_{3},a_{3},a_{2}),
\end{eqnarray*}
where $a_{i}$ are constants.
These can be distinguished from each other by 
neutron scattering experiments. 

The difference in magnetic point groups is also reflected in 
difference in a possible ME effect.
The ME property tensor $\bm{G}$ in the term 
$G_{ij}H_{i}E_{j}$
in the free energy is easily written down for each magnetic 
point group \cite{birss}:
\[
\mbox{(C-1)} 
\left(
\begin{array}{ccc}  0 & G_{1} &0 \\
  -G_{1}&0&0\\
  0&0&0       
\end{array}
\right), \ 
\mbox{(D-1)}
\left(
\begin{array}{ccc}
  0 & G_{1} &0\\
  G_{1}&0& 0\\
  0&0&0       
\end{array}
\right), 
\]
in the original Cartesian coordinates of the cubic lattice. 
The property tensor $\bm{G}$ is
antisymmetric in (C-1), which implies 
$\bm{M}\bot\bm{E}$ and 
$\bm{P}\bot\bm{B}$.
Thus, measurement of the ME effect with a single crystal of 
CaB$_6$ will reveal which of these two cases is realized.
Experimentally, there are two types of AF domains, 
and the sign of the property tensor
$\bm{G}$ is reversed when the staggered magnetization is 
reversed. In measurement of the ME effect, alignment of 
the domain structure is necessary. It is accomplished by means of 
a magnetoelectric annealing, in which the sample is 
cooled under both electric and  magnetic fields.
Experimentally, CaB$_6$ is not insulating, though the sample dependence 
is rather large. Therefore, we cannot apply electric field into the 
sample; we can neither perform magnetoelectric cooling nor
the measurement of magnetoelectric effect.
We should then resort to the optical measurements described in 
the next section.

Domain boundaries between the two AF domains can exhibit interesting 
properties. As in boundaries between two SC domains with 
broken time-reversal symmetry \cite{su}, localized current and 
magnetic moment are induced near the boundary. In the present case 
it is interpreted as the ME effect.

It is reported that weak diamagnetism is often observed in CaB$_{6}$ 
\cite{morikawa,young}. 
It can be attributed to orbital motion, as in the Landau diamagnetism.
In semimetals like Bi and narrow-gap semiconductors, such diamagnetism 
appears.  In Bi, in particular, it is proposed as orbital diamagnetism 
\cite{fukuyama}. Hence, it is no wonder that the small band gap/overlap 
in CaB$_{6}$ originates diamagnetism. 
In reality, it is easily hidden by the AF or by the
ferromagnetism near the surface.

\subsection{Optical Properties}
Optical measurements are known as powerful tools in identifying 
the magnetic properties of the material.
In optical measurements, there are wide range of choices for 
tuning parameters, such as polarization and incident direction 
of light, providing us with lots of information of the material.
One of the most spectacular examples is Cr$_2$O$_3$, where
the second-harmonic generation (SHG) 
is utilized to get the image of the 
antiferromagnetic domains for the first time \cite{ffsp}.
As shown in this example, magnetoelectrics such as Cr$_2$O$_3$
generally exhibit novel optical properties.
Since our theory predicts that the present material 
CaB$_6$ is among the magnetoelectrics, its magnetic symmetry 
can be checked by 
various optical experiments.
For example,  
nonreciprocal (NR) optical effects \cite{bst,fuchs,gr92} 
occur in the magnetoelectrics, since the 
ME tensor appears in the formulation of the NR effects \cite{gr92}.
In other words the ME effect corresponds to a static limit of the
optical NR effects.
Possible NR effects for all the magnetic point groups of 
magnetoelectrics are
calculated in Ref.~\onlinecite{gr92} for transmission and in 
Ref.~\onlinecite{gr99} for reflection.
Let us apply the essential results 
in these references.
Here we note a difference in notation from Ref.~\onlinecite{gr92,gr99};
for (D-1) $4'/{\rm m}'{\rm m}'{\rm m}$ 
the $x$ and $y$ axes are rotated by 45 degrees around the 
$z$-axis from 
those in Ref.~\onlinecite{gr92,gr99}.

\subsubsection{Optical Nonreciprocal Effects}
According to Ref.\onlinecite{gr92}, the multipole expansion of electric and 
magnetic fields gives rise to various tensors, each of which is 
responsible for various optical effects. 
To the order of electric quadrupoles and magnetic dipoles, 
nonvanishing tensors in an $RI$-invariant system such as 
CaB$_6$ are the polar i-tensor 
$\alpha_{\alpha\beta}(\omega)$, 
the polar c-tensor
$a'_{\alpha\beta\gamma}(\omega)(=a'_{\alpha\gamma\beta}(\omega))
$,
and the axial c-tensor $G_{\alpha\beta}(\omega)$ 
in the notation of Ref.~\onlinecite{gr92}, 
where $\alpha_{\alpha\beta}(\omega)$ is the polarizability tensor, while
$a'_{\alpha\beta\gamma}(\omega)$ and $G_{\alpha\beta}(\omega)$ 
are assigned to the 
NR optical effects.
Note that the static limit 
$G_{\alpha\beta}(\omega=0)$ is just the ME property tensor 
$G_{\alpha\beta}$. 
Magnetic symmetry determines nonvanishing components of these tensors. 
We can then calculate the refractive index using these
tensors. The detailed derivation is developed in Appendix
\ref{appendix-b}, and here we present only the results.
Let $n=n'+{\rm i}n''$ denotes the complex refractive index,
where
the real part $n'$ is the refractive index and the 
imaginary part $n''$ is
the absorption 
coefficient.
We also define $\bm{\sigma}$ as $\bm{\sigma}=\bm{k}/k$, i.e. 
a normal vector along the incident direction of light.
The refractive index $n$
for light along the $\pm z$-direction $\bm{\sigma}
=(0,0,\sigma_{z})=(0,0,\pm 1)$
is obtained for two types of polarization as 
\begin{eqnarray*}
&&\mbox{(C-1)}:\\
&&\makebox[5mm]{}
\left[\bm{E}\|\hat{x},\bm{E}\|\hat{y}\right]:
n=\frac{c}{2}\left(
\sigma_{z}
\mu_{0}C_{5}+\sqrt{\mu_{0}^{2}C_{5}^{2}+4\mu_{0}\epsilon_{x}}
\right),\\
&&\mbox{(D-1)}: 
\left[\bm{E}\| \hat{x}
\right]:\ n=\frac{c}{2}\left(
\sigma_{z}
\mu_{0}C_{5}+\sqrt{\mu_{0}^{2}C_{5}^{2}+4\mu_{0}\epsilon_{x}}
\right),\\
&&\makebox[3em]{}
\left[\bm{E}\| \hat{y}\right]:\ n=\frac{c}{2}\left(
-\sigma_{z}
\mu_{0}C_{5}+\sqrt{\mu_{0}^{2}C_{5}^{2}+4\mu_{0}\epsilon_{x}}
\right),
\end{eqnarray*}
where $C_{5}=2G_{xy}+\omega a'_{xxz}$.
Note that since $G_{\alpha\beta}$ and $a'_{\alpha\beta\gamma}$
are c-tensors, $C_{5}$ changes sign for different
AF domains, which are related with each other by the time-reversal.
Thus, for a single domain, (C-1) shows a {\it directional 
birefringence},  but there is no anisotropy in polarization, i.e. 
no birefringence in the usual sense.
In (D-1), on the other hand, directional birefringence and usual
birefringence both exist.
Hence, linearly-polarized light can distinguish between these 
two cases. In fact we do not even need to fix polarization;
unpolarized light is sufficient to distinguish between these
two possibilities. Unpolarized light, which is
an incoherent superposition of both $x$- and $y$-polarized lights, 
shows a refractive index of an average of two refractive indices 
for two types of polarizations 
\begin{eqnarray}
\mbox{(C-1)}&:& n=\frac{c}{2}\left(
-\sigma_{z}
\mu_{0}C_{5}+\sqrt{\mu_{0}^{2}C_{5}^{2}+4\mu_{0}\epsilon_{x}}
\right),\\
\mbox{(D-1)}&:& 
n=\frac{c}{2}\sqrt{\mu_{0}^{2}C_{5}^{2}+4\mu_{0}\epsilon_{x}}.
\end{eqnarray}
Therefore, only in (C-1) does the system show directional 
birefringence for unpolarized light along the tetragonal axis.
This effect, i.e. directional birefringence for unpolarized lights, is 
called a magnetochiral effect \cite{rikken}.
This effect is more prominent in the high-frequency
region
as evidenced in the X-ray measurement of Cr-doped V$_2$O$_3$ \cite{x}.

Experimentally, domain structure is formed, which might obscure the 
experimental results. If we assume that the sample is a single crystal with 
perfect cubic structure above the N{\'e}el temperature $T_{N}$, 
there are two sources of domain formation below $T_{N}$.
One is (a) the AF domains of opposite staggered 
moments, while the tetragonal axis are common.
The other is (b) 
the domains due to different direction of tetragonal distortion below $T_{N}$.
This domain formation will affect the NR optical effects in the
following way.
For the (a)-type domains, the parameters $C_{5}$ has
different signs for different AF domains, which smears out the 
NR effects. Thus, the thin-film sample should have a single domain in
the direction of light propagation.
On the other hand, for the (b)-type domains, we can investigate the 
problem by setting the incident direction as $x$ or $y$ direction in the 
above equations. The result for $\bm{\sigma}=(\sigma_{x},0,0)
=(\pm 1,0,0)$ is 
\begin{eqnarray*}
&&
\left[\bm{E}\| \hat{y}
\right]:\ n=\sqrt{\epsilon_{0}^{-1}\epsilon_{x}},\\
&&
\left[\bm{E}\| \hat{z}\right]:\ n=\sqrt{\epsilon_{0}^{-1}\epsilon_{x}}\left(
1+\frac{\mu_{0}}{\epsilon_{x}}C_{7}
\right)^{-\frac{1}{2}},
\end{eqnarray*}
for both (C-1) and (D-1), where we set
$C_{7}=-G_{xy}+\frac{1}{2}\omega (a'_{zxx}+a'_{xxz})$.
For the incident direction parallel to the $y$-direction, the 
result is similar, with $\bm{E}\|\hat{y}$ 
and $\bm{E}\|\hat{x}$ being interchanged in the above result.
These imply that for $\bm{\sigma}\|\hat{x}$ and $\bm{\sigma}\|\hat{y}$
the NR optical effects do not take place. 
Thus only the light traveling parallel to the $z$-axis undergoes the 
NR optical effect; hence the (b)-type domains will not hinder the 
identification of ground states by the NR effect, although 
it is reduced by the factor of 3.

In reflection, on the other hand, within the method in Ref.~\onlinecite{gr99},
the reflection matrix $R$ for light propagating along the $z$-axis
defined by
\begin{equation}
(E_{{\rm r}})_{j}=R_{jk}
(E_{{\rm i}})_{k}
\end{equation} 
is proportional to identity for both (C-1) and (D-1),
where $E_{\text{i}}$ and $E_{\text{r}}$ are the electric fields for 
incident and reflected lights, respectively.
This implies that there is no 
optical NR effect in reflection normal to the $xy$-plane.
Nevertheless, since this effect is not 
prohibited by symmetry, this NR effect in reflection can emerge
in higher order than in Ref.~\onlinecite{gr99}.
We note that in reflection experiments, the surface
inevitably affects the spectrum \cite{dp}. 
Thus, even if the NR effect in reflection are observed, it might 
be difficult to separate the result into the surface and 
bulk contributions, as it is controversial in Cr$_2$O$_3$.

\subsubsection{Second Harmonic Generation}
Second harmonic generation (SHG) is recently emerging as a new
and powerful tool for determination of magnetic symmetry of 
the material.
This effect appears both in transmission and in reflection. For 
separating bulk contribution from surface one, 
transmission measurement is preferable.
At the surface the inversion symmetry is always broken, and the SHG
always emerges. 
It is difficult to separate bulk and surface contributions in
reflection measurement.

This SHG can arise from various multipole contributions, among 
which an electric-dipole one is the most dominant in general. This
arises from the third 
nonlinear dielectric constant $\chi_{ijk}^{e}$ defined as
\begin{equation}
 P_{i}(2\omega)=\chi_{ijk}^{e}
(2\omega;\omega,\omega) E_{j}(\omega)E_{k}(\omega)
\end{equation}
and is nonzero only if the spatial inversion $I$ is broken. 
Thus this electric-dipole SHG appears either 
at a bulk with broken inversion symmetry or at surfaces.
Depending on magnetic symmetry of the compound, the SHG emerges
in a different way by varying the incident direction or polarization 
of light, and by changing which polarization is measured.
Therefore, by measuring the SHG intensity for various
experimental settings, magnetic symmetry of the material 
can be almost uniquely determined, as the recent examples 
of YMnO$_{3}$ and Cr$_{2}$O$_{3}$ show 
\cite{ffkp,spin-flop,YMO98,YMO99}. 

CaB$_6$ is not an exception.
Two candidates for its magnetic symmetry, (C-1) $4/\text{m}'\text{mm}$ 
and (D-1) $4'/\text{m}'\text{m}'\text{m}$, 
can be distinguished by the SHG, as we shall see below. 
Because CaB$_6$ breaks inversion symmetry according to our prediction, 
this electric-dipole SHG is nonvanishing in the bulk.
The third nonlinear dielectric constant $\chi_{ijk}^{e}$ is decomposed 
into two parts $\chi_{ijk}^{e}=\chi_{ijk}^{e(i)}+\chi_{ijk}^{e(c)}$,
where $\chi_{ijk}^{e(i)}$ and $\chi_{ijk}^{e(c)}$ are $i$-tensors 
(invariant under time-reversal $R$) and $c$-tensors (change sign under 
$R$), respectively.
In CaB$_6$, the $RI$-invariance yields 
$\chi_{ijk}^{e(i)}=0$.
The magnetic point-group symmetry of the system determines  
which components of the tensor $\chi_{ijk}^{e(c)}$ are vanishing.
We can calculate the intensity of the SHG for any choice of incident 
direction and polarization. Among the various choices the simplest one
is the light propagating along the $z$-axis, i.e. 
the $c$-axis in the tetragonal crystal. As shown in Appendix 
\ref{appendix-c}, however, 
the SHG cannot arise in this setting.
Thus, the sample must be tilted to 
observe the SHG, as in YMnO$_3$ \cite{YMO98,YMO99}.
The simplest choice of tilting of the sample 
is around 
the [100] axis. It is, however, not convenient for 
the identification of magnetic symmetry for CaB$_6$
; the difference between the two possibilities (C-1) and 
(D-1) is not conspicuous with this choice.
Instead, we shall tilt the crystal around the [110]
axis while the incident direction is fixed as $\bm{k}\|\hat{z}$. 
Let $\phi$ denote the tilting angle \cite{phi}, 
which is fixed throughout the 
whole measurement.
The details of the calculation is presented in Appendix
\ref{appendix-c}.
Among various choices of polarization of the incoming light 
$\bm{E}(\omega)$ and that measured by the detector, we
find the following; when we wish to see the difference between 
the two cases (C-1) (D-1), the best way is to measure the 
SHG intensity for the polarization perpendicular
to that of the incoming light $\bm{E}(\omega)$.
Let $P(2\omega)_{\bot}$ denote the component 
of $\bm{P}(2\omega)$ perpendicular to the incident polarization. Then
we get 
\begin{eqnarray}
&&\mbox{(C-1)}:P_{\bot}=
E_{x}\frac{\epsilon_{0}\sin\phi}{\sqrt{E_{x}^{2}+E_{y}^{2}}}
\left(
2\chi_{xxz}^{e(c)}E_{y}^{2}\sin^{2}\phi\right.\nonumber \\
&&
\makebox[8mm]{}\left.
-\chi_{zxx}^{e(c)}(E_{x}^{2}+E_{y}^{2}\cos^{2}\phi)
-\chi_{zzz}^{e(c)}E_{y}^{2}\sin^{2}\phi
\right),\label{c-1-pbot}\\
&&\mbox{(D-1)}:P_{\bot}=
E_{y}\frac{\epsilon_{0}\sin\phi}{\sqrt{E_{x}^{2}+E_{y}^{2}}}
\left(
2\chi_{xxz}^{e(c)}(E_{y}^{2}-E_{x}^{2})\cos\phi
\right.\nonumber\\
&&\makebox[8mm]{}\left.-2\chi_{zxx}^{e(c)}E_{x}^{2}\cos\phi
-\chi_{zzz}^{e(c)}E_{x}E_{y}\sin^{2}\phi
\right).\label{d-1-pbot}
\end{eqnarray}
The SHG intensity is proportional to $P_{\bot}^{2}$.
Thus we propose the following experiment for distinguishing (C-1)
and (D-1).
As the polarization of the incoming 
light is rotated, and polarization measured by 
the detector is rotated accordingly as is perpendicular 
to $\bm{E}(\omega)$, we should measure 
the change of the SHG intensity. In 
(C-1) the SHG intensity vanishes when $E_{x}=0$ and 
in (D-1) it vanishes when $E_{y}=0$. Thus, in this 
measurement we can distinguish between these two.
This SHG is observed only below $T_{N}$, which will be an evidence of 
magnetic ordering in the parent compound CaB$_6$.

Let us consider an effect of domain formation on the SHG. 
For the (a)-type domains, $\chi$'s change sign for different 
domains; this does not, however, affect the SHG intensity,
which is proportional to $P_{\bot}^{2}$. Thus the type (a) domains 
will do no harm. On the other hand, 
the (b)-type domains will kill the difference between
(C-1) and (D-1), as explained below. 
If the tetragonal distortion is along the $x$-
or $y$-axis instead of the $z$-axis, we can calculate in the similar 
way as above. The resulting form of $P_{\bot}$ is so complicated that 
we do not reproduce here. Unlike (\ref{c-1-pbot})
(\ref{d-1-pbot}), $P_{\bot}$ cannot be factorized, i.e. there is
no node when the incident polarization is rotated. If this 
type of domains are mixed along the direction of light propagation, the
SHG intensity does not have no node, and we can no longer 
distinguish  between (C-1) and (D-1). Therefore, we should make sure 
that this (b)-type domains are not mixed. This can be 
verified by an untilted $(\phi=0)$ setting. If there is only a 
single domain, we can use (\ref{c-1-pbot})(\ref{d-1-pbot})
and the SHG intensity vanishes when $\phi=0$. An existence
of the (b)-type domains will give nonzero intensity.
To avoid the mixture of these (b)-type domains, a choice of substrate 
material will be important. It is expected that a good choice of substrate
material for thin-film fabrication 
will uniquely set the direction of tetragonal distortion to be 
perpendicular to the thin-film.

In cases such as Cr$_2$O$_3$, contribution from magnetic dipole cannot 
be neglected. Let us, therefore, consider this contribution for CaB$_{6}$. 
This comes from the SH 
magnetic-dipole tensor $\chi_{ijk}^{m}$ defined by
\begin{equation}
M_{i}(2\omega)=\frac{\epsilon_{0}c}{n}
\chi_{ijk}^{m}(2\omega;\omega,\omega)E_{j}(\omega)
E_{k}(\omega).
\end{equation}
Since this is an axial tensor, the $RI$-symmetry in CaB$_6$ allows 
only the $i$-tensor component $\chi_{ijk}^{m(i)}$.
Hence it persists both below and above the N{\'e}el temperature. 
Nevertheless, we show in Appendix \ref{appendix-d}
that when we measure the polarization of the SHG 
perpendicular to $\bm{E}(\omega)$, this magnetic-dipole contribution 
never appears. This is another merit of our experimental setting.

The AF domain topography by the SHG \cite{ffsp,spin-flop,YMO00}
is also possible in this material.
Domain topography utilizes an interference between the $c$-tensor 
$\chi^{(c)}$ 
and the $i$-tensor component $\chi^{(i)}$. Thus we should have the SHG
from both contributions at the same time for domain topography. 
At least the detected polarization should not 
be perpendicular to the incident polarization. To obtain better 
contrast for two types of domains, the polarization of the incident
light and that of detected light should be tuned \cite{YMO99}. 

\section{Discussion and Conclusion}
Now we discuss the relevance of our results to various experiments
on CaB$_6$ and Ca$_{1-x}$La$_{x}$B$_6$. We believe many of the novel 
magnetic properties can be interpreted as the ME effect.
The ferromagnetism in the thin-film CaB$_6$ is interpreted as caused by an 
electric field between vacuum and the substrate. 
For the powder sample experiment and the La-doping experiment, 
the explanation is more delicate. We believe that the carriers by La-doping 
is trapped by impurities/defects and create local electric fields.  
Therefore in these cases, an internal electric field 
and/or a gradient of a strain has a random direction, 
and hence the magnetic moment is induced locally due to this mechanism. 
Without an external magnetic field, they almost cancel with each other,
giving zero or quite small uniform magnetization, which appears to 
contradict with the experiments. 
In fact it is consistent with experiments. 
The compound shows hysteresis, which is thought as an experimental
evidence for ``ferromagnetism''.
This is quite opposite to our intuition. The present compound is AF, 
which usually does not show hysteresis behavior in the $M$-$H$ curve; 
it is a novel feature of 
doped AF magnetoelectrics. Hysteresis implies that the free energy 
has a double minimum as a function of magnetization.
In a first sight it is unique to ferromagnets; we propose that 
this is also the case for Ca$_{1-x}$La$_{x}$B$_{6}$, which we
claim is not 
ferromagnet in the bulk.  In this material, there is a local 
electric field induced by doped impurities. 
When we vary the magnetic field, the two AF domains 
will switch to each other to minimize the free energy. 
As we explain below, 
Gibbs free energy acquires
two minima as a function of $M$ in the presence of local electric field, 
which causes hysteresis.
Considering that the ME tensor $G$ is proportional to the 
exciton order parameters $\text{Im}\eta$,
we can write the Gibbs free energy $F$ as 
\begin{eqnarray}
&&F=-C(\text{Im}\eta) P M+\frac{1}{2\chi}M^{2}+\frac{1}{2\alpha}P^{2}
-A(\text{Im} \eta)^{2}
\nonumber \\
&&\makebox[1cm]{}+B(\text{Im} \eta)^{4}-DP\text{Re}\eta,
\end{eqnarray}
where $A,B,C,\chi,\alpha$ are positive constants and $D$ is a constant. 
Here, for simplicity, 
we omitted subscripts for $P$, $M$, $\eta$ without loss of generality.
By minimizing $F$ in terms of $\text{Im} \eta$, we obtain $\text{Im} \eta\sim 
\text{sgn}(PM)\sqrt{A/2B}$, provided $C$ is small. When we substitute 
it into $F$, we get
\begin{equation}
F\sim -C\sqrt{\frac{A}{2B}}
|P||M|+\frac{1}{2\chi}M^{2}+\frac{1}{2\alpha}P^{2}
-DP\text{Re}\eta.
\end{equation}
Near the doped impurities, the polarization $P$ is induced by a local 
electric field. In this case, $P$, $M$, and $\eta$ has spatial 
dependence
and should be written as $P(\bm{x})$, $M(\bm{x})$, $\eta(\bm{x})$.
Although the sign of $P(\bm{x})$ changes spatially, one can 
basically regard $|P(\bm{x})|$ as a constant.
In the presence of $P$, this Gibbs free energy has two minima as a
function of $M$, implying hysteresis. 
Note that this is not a genuine ferromagnet; 
a uniform $M(\bm{x})$ does not emerge in the absence of an 
external magnetic field, because the uniform $M(\bm{x})$ costs
an elastic energy $(\nabla \text{Im} \eta(\bm{x}))^{2}$.
On the other hand,
If there is no impuritiy, i.e. $P=0$,
$F$ does not have double-minimum structure, and no hysteresis results.
Hence, hysteresis in Ca$_{1-x}$La$_{x}$B$_{6}$ can result from 
the ME effect together with
the local electric field near doped impurities.

Because Ca$_{1-x}$La$_{x}$B$_{6}$ is not insulating, the electric field
is screened. The Thomas-Fermi screening length is estimated as 
\begin{equation}
\frac{\sqrt{a_{0}\epsilon}}{2}\left(\frac{\pi}{3n_{0}}\right)^{\frac{1}{6}}
\sim 10\text{\AA},
\end{equation}
where $a_{0}$ is the Bohr radius, and we used the values
$\epsilon\sim 6 $ for the dielectric 
constant \cite{gavilano} and $n_{0}\sim (1\mbox{-}4)\times 10^{24}
{\rm m}^{-3}$ for a density of electrons \cite{gianno}.
The appearance of magnetic moments via the ME effect is therefore
localized within this 
screening length of the impurities or surfaces.

Other peculiarities of Ca$_{1-x}$La$_x$B$_6$ can also be explained as well.
The high Curie temperature $(\sim 600{\rm K})$ is nothing but 
a N{\'e}el temerature of the parent compound CaB$_6$, and is not
contradictory 
with a tiny magnetic moment. A rather narrow range ($x\lesssim 0.01$) 
of La-doping 
allowing ferromagnetism is attributed to fragility of excitonic order 
by a small amount of impurities \cite{sk,z}.
Moreover, our scenario is also consistent with the experimental results
that deficiency in Ca sites \cite{morikawa} or
doping of divalent elements like Ba \cite{young} or Sr \cite{ott} 
induces ferromagnetism.
It is also confirmed numerically by a supercell approach that
imperfections and surfaces can induce local moments \cite{md}.
It is hard to explain them within the spin-doping scenario \cite{zra}.
Furthermore, strangely enough,
it is experimentally 
hard to find a correlation between magnetism and electrical 
resistivity, as seen in magnetizaion \cite{morikawa} and in 
nuclear magnetic resonance \cite{gavilano}. This novelty can be
considered as natural consequence of our scenario; electrical 
resistivity should be mainly due to doped carriers by La, while the 
magnetization is due to local lattice distortion and/or electric field.

The ESR experiments by Kunii \cite{kunii}
also support the above scenario. The ESR data show that in 
a disk-shaped Ca$_{1-x}$La$_x$B$_6$ ($x=0.005$),
the magnetic moment only exists within the surface layer.
Furthermore, the moment $\bm{M}$ does not orient in the direction
of $\bm{H}$, i.e. it feels strong magnetic anisotropy to keep the 
moment within the disk plane. This might be due to the long-range 
dipolar energy, and not due to the above scenario.
Nevertheless, it is unlikely that the long-range dipolar energy 
causes such a strong anisotropy.
This point requires further experimental and theoretical investigation.
Let us, for the moment, assume that this strong anisotropy 
is mainly caused by the exciton condensation and the ME mechanism.
Since this electric field should be
perpendicular to the plane, the strong easy-plane anisotropy 
parallel to the surface implies that $\bm{M}\bot \bm{E}$.
Therefore, among the four cases, (C-1) or (C-2) are compatible,
i.e. the excitons corresponding to 
$\bm{\eta}(\Gamma_{15},1)$ and 
$\bm{\eta}(\Gamma_{15},2)$ simultaneously condense.
Considering the Raman scattering data, we conclude that 
(C-1) is the only possibility 
compatible with the experiments.

We briefly mention the relationship between our theory and the 
theories in Refs.~\onlinecite{zra,bv}.
In the absence of doping, our theory is consistent with 
Refs.~\onlinecite{zra,bv}. The difference lies in the 
mechanism of ferromagnetism in doping. We proposed that 
defects, impurities or surfaces induce ferromagnetism. This is different
from Refs.~\onlinecite{zra,bv}, in which ferromagnetism is due to 
spin alignment of doped carriers. We should note that these two mechanisms
can coexist. We can distinguish these two contributions for 
ferromagnetism by systematically changing the valence of doped impurities.
In doping with divalent doping, no carriers are doped, and the moment is
due to our scenario.
Difference between trivalent- and divalent-doping corresponds 
to the scenarios in Refs.~\onlinecite{zra,bv}.
Our theory is treating the dilute doping limit.
We also note that because our theory is based on the GL theory,
it is treating an instability in slight doping. Thus our conclusions
do not hinder an appearance of rich phase diagrams, as presented,
for example, in Ref.~\onlinecite{zra,bv,b,vb,bbm}.

We mention here a role of the spin-orbit coupling. 
In the absence of 
the spin-orbit coupling, 
the GL free energy $\Phi^{(2)}$
for the imaginary parts of the order parameters is written as 
\begin{eqnarray*}
 &&\Phi^{(2)}=
\lambda_{+}^{{\rm Im}}
\sum_{i}({\rm Im}\ \eta_{i}(\Gamma_{15},+))^{2}
 \nonumber \\
&&\makebox[.5mm]{}
+\lambda_{-}^{{\rm Im}}
\sum_{i}({\rm Im}\ \eta_{i}(\Gamma_{15},-))^{2}
+\lambda_{\Gamma_{25}}^{{\rm Im}}
\sum_{j}({\rm Im}\ \eta_{j}(\Gamma_{25}))^{2},
\end{eqnarray*}
from (\ref{zero-spin-orbit}) in Appendix \ref{appendix-a}.
As $\lambda_{+}^{{\rm Im}}>\lambda_{-}^{{\rm Im}}=
\lambda_{\Gamma_{25}}^{{\rm Im}}$ from the aforementioned 
microscopic calculation, the condensation of excitons 
in $\Gamma_{15}$ and those in $\Gamma_{25}$ occur simultaneously.
Furthermore,
inspection of $\Phi^{(4)}$ in the absence of the spin-orbit coupling 
shows that the quartic-order terms in the GL free energy do not lift this
degeneracy. 
Sixth-order terms will lift it, and a resulting state
will belong to either 4/${\rm m}'{\rm m}'{\rm m}'$ or  
$4'/{\rm m}'{\rm m}'{\rm m}$. 
The details are presented in Appendix \ref{appendix-a}.
Both of them still lead to the ME 
effect in the absence of the 
spin-orbit coupling;
this ME effect must be generated from an orbital motion.
Thus, the AF state in CaB$_6$ has an orbital nature as 
well as a spin nature. With spin-orbit interaction, these two
are inseparably mixed together.

Recently, similar novel ferromagnets such as CaB$_2$C$_2$ \cite{akimitsu}
and rhombohedral C$_{60}$ \cite{makarova} have been discovered. 
They share several properties 
with CaB$_6$, i.e. high Curie temperature, smallness of the moment and 
lack of partially-filled $d$- or $f$-bands. They might
be explained by the similar scenario as in CaB$_6$, and indeed one of 
the authors explained the novel ferromagnetism in CaB$_2$C$_2$ within the 
scenario of exciton condensation \cite{murakami}.

In conclusion, we have studied the symmetry properties of the 
excitonic state in the parent compound CaB$_6$, and found that the 
triplet excitonic state with broken time-reversal and inversion symmetries 
offers a natural explanation, in terms of the ME effect, 
for the novel ferromagnetism
emerging in La-doping or thin-film fabrication. 
This scenario can be tested experimentally by
measurements of the ME effect and
the optical non-reciprocal effect in single crystal of the parent compound
CaB$_6$.

\begin{acknowledgments}
The authors thank helpful discussion with Y.~Tokura, H. Takagi,
Y.~Tanabe,
J.~Akimitsu, M.~Udagawa,
and K.~Ohgushi. We acknowledge support by 
Grant-in-Aids
from the Ministry of Education, Culture, Sports, Science and Technology 
and RFBR Grant No. 01-02-16508.
\end{acknowledgments}

\appendix
\section{Derivation of the refractive index -- optical nonreciprocal
 effects --}
\label{appendix-b}
Nonvanishing components for each tensor of the optical nonreciprocal 
(NR) effects are determined from each 
magnetic point group as \cite{birss}
\begin{eqnarray*}
\mbox{(C-1)}&:&\alpha_{xx}=\alpha_{yy},\ \ \ \alpha_{zz},\\
&&G_{xy}=-G_{yx},\\
&&a'_{xxz}=a'_{xzx}=a'_{yyz}=a'_{yzy},\\
&&  a'_{zxx}=a'_{zyy},\ \ \ a'_{zzz},\\
\mbox{(D-1)}&:&\alpha_{xx}=\alpha_{yy},\ \ \ \alpha_{zz},\\
&&G_{xy}=G_{yx},\\
&&a'_{xxz}=a'_{xzx}=-a'_{yyz}=-a'_{yzy},\\
&&  a'_{zxx}=-a'_{zyy}.
\end{eqnarray*}
In calculating the optical properties in transmission or in 
reflection, the tensor $\tilde{A}_{\alpha\beta\gamma}$
defined as
\begin{eqnarray}
&&\tilde{A}_{\alpha\beta\gamma}
=\tilde{A}_{\beta\alpha\gamma}\nonumber \\
&&\ \ =-\epsilon_{\beta\gamma\delta}G_{\alpha\delta}
-\epsilon_{\alpha\gamma\delta}G_{\beta\delta}
+\frac{1}{2}\omega(a'_{\alpha\beta\gamma}+
a'_{\beta\alpha\gamma}
)
\end{eqnarray}
plays a central role \cite{gr92}.
Its nonvanishing components for each magnetic group
are summarized in Table 2 of Ref.~\onlinecite{gr92}. They are 
\begin{eqnarray*}
\mbox{(C-1)}&:&\tilde{A}_{zxx}=
\tilde{A}_{xzx}=
\tilde{A}_{zyy}=
\tilde{A}_{yzy}\\
&&\makebox[1cm]{}
=-G_{xy}+\frac{1}{2}\omega(a'_{zxx}+a'_{xxz}),\\
&&\tilde{A}_{xxz}=
\tilde{A}_{yyz}=2G_{xy}+\omega a'_{xxz},\\
&&
\tilde{A}_{zzz}=\omega a'_{zzz},\\
\mbox{(D-1)}&:&\tilde{A}_{zxx}=
\tilde{A}_{xzx}=
-\tilde{A}_{zyy}=
-\tilde{A}_{yzy}\\
&&\makebox[1cm]{}
=-G_{xy}+\frac{1}{2}\omega(a'_{zxx}+a'_{xxz}),\\
&&\tilde{A}_{xxz}=
-\tilde{A}_{yyz}=2G_{xy}+\omega a'_{xxz},
\end{eqnarray*}
The refractive index $n$ for each polarization is obtained as a solution 
of the following equation.
\begin{eqnarray}
&&\left[
n^{2}(\sigma_{\alpha}\sigma_{\beta}-\delta_{\alpha\beta})+\delta_{\alpha\beta}
+\epsilon_{0}^{-1}\alpha_{\alpha\beta}+c\mu_{0} n
\sigma_{\gamma}
\tilde{A}_{\alpha\beta\gamma}\right]\nonumber \\
&&\cdot E_{\beta}=0.
\end{eqnarray}
For incident direction along the $\pm z$-direction $\bm{\sigma}
=\bm{k}/k=(0,0,\sigma_{z})=(0,0,\pm 1)$, 
an equation determining the refractive 
index $n$ is 
\begin{widetext}
\begin{eqnarray}
&&\mbox{(C-1)}:\left(
\begin{array}{ccc}
-n^{2}+\epsilon_{0}^{-1}\epsilon_{x}+c\mu_{0}nC_{5}\sigma_{z}&&\\
&-n^{2}+\epsilon_{0}^{-1}\epsilon_{x}+c\mu_{0}nC_{5}\sigma_{z}&\\
&&\epsilon_{0}^{-1}\epsilon_{z}+c\mu_{0}nC_{6}\sigma_{z}
\end{array}
\right)
\left(\begin{array}{c}
E_{x}\\E_{y}\\E_{z}
\end{array}
\right)=0,\\
&&\mbox{(D-1)}:\left(
\begin{array}{ccc}
-n^{2}+\epsilon_{0}^{-1}\epsilon_{x}+c\mu_{0}nC_{5}\sigma_{z}&&\\
&-n^{2}+\epsilon_{0}^{-1}\epsilon_{x}-c\mu_{0}nC_{5}\sigma_{z}&\\
&&\epsilon_{0}^{-1}\epsilon_{z}
\end{array}
\right)
\left(\begin{array}{c}
E_{x}\\E_{y}\\E_{z}
\end{array}
\right)=0,
\end{eqnarray}
\end{widetext}
where $\epsilon_{x}=\epsilon_{0}+\alpha_{xx}=\epsilon_{0}+\alpha_{yy}$,
$\epsilon_{z}=\epsilon_{0}+\alpha_{zz}$
$C_{5}=2G_{xy}+\omega a'_{xxz}$, and $C_{6}=\omega a'_{zzz}$.
Thus for both cases the refractive index $n$
is given by 
\begin{eqnarray*}
&&\mbox{(C-1)}: \left[\bm{E}\|\hat{x},\ 
\bm{E}\|\hat{y}\right]:
n=\frac{c}{2}\left(
\sigma_{z}\mu_{0}C_{5}+\sqrt{\mu_{0}^{2}C_{5}^{2}+4\mu_{0}\epsilon_{x}}
\right),\\
&&\mbox{(D-1)}: 
\left[\bm{E}\| \hat{x}
\right]:\ n=\frac{c}{2}\left(
\sigma_{z}\mu_{0}C_{5}+\sqrt{\mu_{0}^{2}C_{5}^{2}+4\mu_{0}\epsilon_{x}}
\right),\\
&&\makebox[3em]{}
\left[\bm{E}\| \hat{y}\right]:\ n=\frac{c}{2}\left(
-\sigma_{z}\mu_{0}C_{5}+\sqrt{\mu_{0}^{2}C_{5}^{2}+4\mu_{0}\epsilon_{x}}
\right),
\end{eqnarray*}
for $\bm{\sigma}=(0,0,\pm 1)$.

In order to consider an effect of domain formation of the (b)-type, 
i.e. domains with different directions of distortion, let us
study the light propagating along the $x$-direction in the similar way
as above.
An equation for $\bm{\sigma}=(\sigma_{x},0,0)
=(\pm 1,0,0)$ is 
\[
\left(
\begin{array}{ccc}
\epsilon_{0}^{-1}\epsilon_{x}&&c\mu_{0}nC_{7}\sigma_{x}\\
&-n^{2}+\epsilon_{0}^{-1}\epsilon_{x}&\\
c\mu_{0}nC_{7}\sigma_{x}
&&-n^{2}+\epsilon_{0}^{-1}\epsilon_{z}
\end{array}
\right)
\left(\begin{array}{c}
E_{x}\\E_{y}\\E_{z}
\end{array}
\right)=0, 
\]
for both (C-1) and (D-1), where we set
$C_{7}=-G_{xy}+\frac{1}{2}\omega (a'_{zxx}+
a'_{xxz})$. Thus the refractive index is 
\begin{eqnarray*}
&&
\left[\bm{E}\| \hat{y}
\right]:\ n=\sqrt{\epsilon_{0}^{-1}\epsilon_{x}},\\
&&
\left[\bm{E}\| \hat{z}\right]:\ n=\sqrt{\epsilon_{0}^{-1}\epsilon_{z}}\left(
1+\frac{\mu_{0}}{\epsilon_{x}}C_{7}
\right)^{-\frac{1}{2}},
\end{eqnarray*}
for $\bm{\sigma}=(\pm 1,0,0)$.
The latter one with $\bm{E}\|\hat{z}$ contains
a longitudinal component of $\bm{E}$, i.e. parallel to $\hat{x}$.
It is called an S-wave (skew wave)
in Ref.~\onlinecite{gr92}, and can only propagate only 
inside the sample.
For $\bm{\sigma}=(0,\pm 1,0)$, we get 
\begin{eqnarray*}
&&
\left[\bm{E}\| \hat{x}
\right]:\ n=\sqrt{\epsilon_{0}^{-1}\epsilon_{x}},\\
&&
\left[\bm{E}\| \hat{z}\right]:\ n=\sqrt{
\epsilon_{0}^{-1}\epsilon_{z}}\left(
1+\frac{\mu_{0}}{\epsilon_{x}}C_{8}
\right)^{-\frac{1}{2}},
\end{eqnarray*}
These imply that for $\bm{\sigma}\|\hat{x}$ and $\bm{\sigma}\|\hat{y}$
the NR optical effects do not take place. 

\section{Calculation of the SHG intensity to distinguish (C-1) and (D-1)}
\label{appendix-c}
The nonvanishing components of the 
nonlinear electric-dipole tensor 
$\chi_{ijk}^{e(c)}$ (polar c-tensor) are given as \cite{birss}
\begin{eqnarray*}
\mbox{(C-1)}&:& 
\chi_{zzz}^{e(c)},\ \ \ \chi_{zxx}^{e(c)}=\chi_{zyy}^{e(c)},\\
&&
\chi_{xzx}^{e(c)}=\chi_{xxz}^{e(c)}=
\chi_{yzy}^{e(c)}=\chi_{yyz}^{e(c)},\\
\mbox{(D-1)}&:& \chi_{zzz}^{e(c)},\ \ \  
\chi_{zxx}^{e(c)}=-\chi_{zyy}^{e(c)},\\ 
&& \chi_{xzx}^{e(c)}=\chi_{xxz}^{e(c)}=
-\chi_{yzy}^{e(c)}=-\chi_{yyz}^{e(c)}.
\end{eqnarray*}
By using them, we can derive the equations for the SHG, as in 
Cr$_2$O$_3$\cite{ffkp}. 
The nonlinear polarization  $\bm{P}(2\omega)$ induced by 
$\chi_{ijk}^{e}$ is 
\begin{eqnarray}
&&\mbox{(C-1)}: \bm{P}(2\omega) =\epsilon_{0}
\left(
\begin{array}{c}
 2\chi_{xxz}^{e(c)}E_{x}E_{z}\\
 2\chi_{xxz}^{e(c)}E_{y}E_{z}\\
\chi_{zxx}^{e(c)}
(E_{x}^{2}+E_{y}^{2})
+\chi_{zzz}^{e(c)}
E_{z}^{2}
\end{array}
\right)\label{PNL-C-1}
\\
&&\mbox{(D-1)}: \bm{P}(2\omega) =
\epsilon_{0}
\left(
\begin{array}{c}
 2\chi_{xxz}^{e(c)}E_{x}E_{z}\\
-2\chi_{xxz}^{e(c)}E_{y}E_{z}\\
\chi_{zxx}^{e(c)}
(E_{x}^{2}-E_{y}^{2})
+\chi_{zzz}^{e(c)}
E_{z}^{2}
\end{array}
\right )\label{PNL-D-1}
\end{eqnarray}
These contribute to the source term $\bm{S}^{e}$ for the SHG \cite{shen}
as 
\[
 \bm{S}^{e}
=\mu_{0}\frac{\partial^{2}\bm{P}(2\omega)}{\partial t^{2}}
=-4\omega^{2}\mu_{0}\bm{P}(2\omega).
\]
Therefore, light propagating along the $z$-axis, i.e. 
the $c$-axis in the tetragonal crystal, $\bm{S}$ is 
along the $z$-axis and the SHG cannot be generated.
Thus, the incident direction must be tilted to 
observe the SHG, as in YMnO$_3$ \cite{YMO98,YMO99}.
For practical calculations, it is more convenient to tilt the crystal while 
fixing the light propagation 
along the $z$-axis. This kind of treatment is also 
adopted in Ref.~\onlinecite{YMO99}.
The simplest choice of tilting of the sample 
is around 
the [100] axis, but this choice does not 
manifest clearly the difference between (C-1) and 
(D-1).
Instead, we shall tilt the crystal around the [110]
axis while the incident direction is fixed as $\bm{k}\|\hat{z}$. 
We fix the tilting angle $\phi$, and we fix its value throughout the 
whole measurement.
The calculation is similar to the one for YMnO$_3$
in Ref.~\onlinecite{YMO99}. 
The new coordinate system ($A'_{x},A'_{y},A'_{z}$) 
fixed to the crystal is related to the original one 
($A_{x},A_{y},A_{z}$) as follows;
\begin{equation}
\left(\begin{array}{c}
A'_{x}\\A'_{y}\\A'_{z}
\end{array}
\right)
=
\left(\begin{array}{ccc}
\frac{1}{\sqrt{2}} &
\frac{1}{\sqrt{2}} &\\
-\frac{1}{\sqrt{2}} &
\frac{1}{\sqrt{2}} &\\
&&1\\
\end{array}
\right)
\left(\begin{array}{ccc}
1&&\\&\cos\phi&\sin\phi\\
&-\sin \phi&\cos\phi
\end{array}
\right)
\left(\begin{array}{c}
A_{x}\\A_{y}\\A_{z}
\end{array}
\right).
\end{equation}
First, the polarization $\bm{E}(\omega)=(E_{x},E_{y},0)$ is 
transformed to the new coordinate. Then (\ref{PNL-C-1})
(\ref{PNL-D-1}) are applied to get the nonlinear 
polarization $\bm{P}(2\omega)$, and transform it back 
to the original coordinate ($A_{x},A_{y},A_{z}$). 
The resulting form for the nonlinear
polarization is
\begin{widetext}
\begin{eqnarray}
&&\mbox{(C-1)}:\bm{P}(2\omega)=
\epsilon_{0}
\left(
\begin{array}{c}
-2\chi_{xxz}^{e(c)}E_{x}E_{y}\sin\phi\\
-2\chi_{xxz}^{e(c)}E_{y}^{2}\cos^{2}\phi\sin\phi
-\chi_{zxx}^{e(c)}(E_{x}^{2}+E_{y}^{2}\cos^{2}\phi)
\sin\phi
-\chi_{zzz}^{e(c)}E_{y}^{2}\sin^{3}\phi\\
-2\chi_{xxz}^{e(c)}E_{y}^{2}\cos\phi\sin^{2}\phi
+\chi_{zxx}^{e(c)}(E_{x}^{2}+E_{y}^{2}\cos^{2}\phi)
\cos\phi
+\chi_{zzz}^{e(c)}E_{y}^{2}\cos\phi\sin^{2}\phi\\
\end{array}
\right),\\
&&\mbox{(D-1)}:\bm{P}(2\omega)=
\epsilon_{0}
\left(
\begin{array}{c}
-2\chi_{xxz}^{e(c)}E_{y}^{2}\cos\phi\sin\phi\\
-2(\chi_{xxz}^{e(c)}+\chi_{zxx}^{e(c)})
E_{x}E_{y}\cos\phi\sin\phi
-\chi_{zzz}^{e(c)}E_{y}^{2}\sin^{3}\phi\\
-2\chi_{xxz}^{e(c)}E_{x}E_{y}\sin^{2}\phi
+2\chi_{zxx}^{e(c)}
E_{x}E_{y}\cos^{2}\phi
+\chi_{zzz}^{e(c)}E_{y}^{2}\sin^{2}\phi\cos\phi
\end{array}
\right).
\end{eqnarray}
\end{widetext}
In order to see the difference between 
(C-1) and (D-1), we find that the best way is to measure the 
SHG intensity for the polarization perpendicular
to that of the incoming light $\bm{E}(\omega)$.
The projection of $\bm{P}(2\omega)$ onto the 
direction perpendicular to both $\bm{E}(\omega)$ and 
$z$-axis is just $P(2\omega)_{\bot}=\bm{P}
(2\omega)\cdot \bm{n}_{\bot}$, where 
$\bm{n}_{\bot}=E_{0}^{-1}(-E_{y},\ E_{x},\ 0), \ E_{0}=
\sqrt{E_{x}^{2}+E_{y}^{2}}$. Thus
\begin{eqnarray}
\mbox{(C-1)}&:&P_{\bot}=
\frac{\epsilon_{0}}{\sqrt{E_{x}^{2}+E_{y}^{2}}}E_{x}
\sin\phi\left(
2\chi_{xxz}^{e(c)}E_{y}^{2}\sin^{2}\phi\right.\nonumber \\
&&
\makebox[1em]{}\left.
-\chi_{zxx}^{e(c)}(E_{x}^{2}+E_{y}^{2}\cos^{2}\phi)
-\chi_{zzz}^{e(c)}E_{y}^{2}\sin^{2}\phi
\right),\label{c-1-pbot-2}\\
\mbox{(D-1)}&:&P_{\bot}=
\frac{\epsilon_{0}}{\sqrt{E_{x}^{2}+E_{y}^{2}}}E_{y}\sin\phi
\left(
2\chi_{xxz}^{e(c)}(E_{y}^{2}-E_{x}^{2})\cos\phi
\right.\nonumber\\
&&\makebox[1em]{}\left.-2\chi_{zxx}^{e(c)}E_{x}^{2}\cos\phi
-\chi_{zzz}^{e(c)}E_{x}E_{y}\sin^{2}\phi
\right).\label{d-1-pbot-2}
\end{eqnarray}
which are identical with (\ref{c-1-pbot})(\ref{d-1-pbot}).
Note that (\ref{c-1-pbot-2}) is proportional to $E_{x}$, while 
(\ref{d-1-pbot-2}) is proportional to $E_{y}$;
this enables an identification of the true ground state.

\section{Magnetic dipole contribution to the SHG}
\label{appendix-d}
Its nonvanishing components are given as
\begin{eqnarray*}
\mbox{(C-1)(D-1)}&:& \chi_{xyz}^{m(i)}=\chi_{xzy}^{m(i)}=-\chi_{yxz}^{m(i)}
=-\chi_{yzx}^{m(i)},\\
&&\chi_{zxy}^{m(i)}
=-\chi_{zyx}^{m(i)}.
\end{eqnarray*}
By using them, we can derive the equations for the SHG, as in 
Cr$_2$O$_3$\cite{ffkp}. 
The nonlinear magnetization $\bm{M}(2\omega)$ induced by 
$\chi_{ijk}^{m}$ is 
\[
\bm{M}(2\omega) =
\frac{2\epsilon_{0}c}{n}\chi_{xyz}^{m(i)}
E_{y}\sin\phi
\left(\begin{array}{c}
-E_{y}\cos\phi\\
E_{x}\cos\phi\\
E_{x}\sin\phi
\end{array}
\right)
\]
These contribute to the source term $\bm{S}^{m}$ 
for the SHG \cite{shen}
as 
\begin{eqnarray*}
 \bm{S}^{m}&=&\mu_{0}\nabla\times\frac{\partial\bm{M}(2\omega)}{
\partial t}\\
&=&
-\frac{8\omega^{2}}{c^{2}}\chi_{xyz}^{m(i)}
E_{y}\sin\phi\cos\phi
\left(\begin{array}{c}
E_{x}\\
E_{y}\\
0
\end{array}
\right)\\
&=&
-\frac{8\omega^{2}}{c^{2}}\chi_{xyz}^{m(i)}
E_{y}\sin\phi\cos\phi\bm{E}.
\end{eqnarray*}
Therefore, this $\bm{S}^{m}$ is parallel to the incident polarization 
$\bm{E}(\omega)$.
Thus, when we measure the polarization of the SHG 
perpendicular to $\bm{E}(\omega)$, this magnetic-dipole contribution 
never appears. 

\section{Limit of Zero Spin-Orbit Coupling}
\label{appendix-a}
In the absence of 
the spin-orbit coupling, 
the GL free energy should be invariant under the spin rotation;
by lengthy calculations this leads to relations
\begin{eqnarray}
&& A_{2}+A_{4}=2A_{1},\ \  A_{2}-A_{4}=A_{3}/\sqrt{2},\nonumber \\
&& B_{2}+B_{4}=2B_{1},\ \  B_{2}-B_{4}=B_{3}/\sqrt{2}.
\label{no-spin-orbit}
\end{eqnarray}
When we substitute them into $\Phi^{(2)}$ we get 
\begin{eqnarray}
 &&\Phi^{(2)}=
\lambda_{+}^{{\rm Im}}
\sum_{i}({\rm Im}\ \eta_{i}(\Gamma_{15},+))^{2}
+\lambda_{-}^{{\rm Im}}
\sum_{i}({\rm Im}\ \eta_{i}(\Gamma_{15},-))^{2}
 \nonumber \\
&&\makebox[.2mm]{}
+\lambda_{\Gamma_{25}}^{{\rm Im}}
\sum_{j}({\rm Im}\ \eta_{j}(\Gamma_{25}))^{2}\nonumber \\
&&
\makebox[.2mm]{
}+\lambda_{+}^{{\rm Re}}
\sum_{i}({\rm Re}\ \eta_{i}(\Gamma_{15},+))^{2}
+\lambda_{-}^{{\rm Re}}
\sum_{i}({\rm Re}\ \eta_{i}(\Gamma_{15},-))^{2}
 \nonumber \\
&&\makebox[.2mm]{}
+\lambda_{\Gamma_{25}}^{{\rm Re}}
\sum_{j}({\rm Re}\ \eta_{j}(\Gamma_{25}))^{2}
\label{zero-spin-orbit}
\end{eqnarray}
where $\bm{\eta}(\Gamma_{15},\pm)$ 
are new basis functions,
defined in (\ref{gamma+}) (\ref{gamma-}).
The coefficients in (\ref{zero-spin-orbit}) are
\begin{eqnarray}
&& \lambda_{+}^{{\rm Im}}=\frac{3B_{2}-B_{4}}{2},\ \ 
\lambda_{-}^{{\rm Im}}=\lambda_{\Gamma_{25}}^{{\rm Im}}=B_{4},\\
&& \lambda_{+}^{{\rm Re}}=\frac{3A_{2}-A_{4}}{2},\ \ 
\lambda_{-}^{{\rm Re}}=\lambda_{\Gamma_{25}}^{{\rm Re}}=A_{4}.
\end{eqnarray}

Eqn.(\ref{zero-spin-orbit}) is invariant under the cubic group
operations in the orbital space. Thus, each term in
(\ref{zero-spin-orbit})
can be classified into irreducible representation of the 
${\rm O}_{{\rm h}}$
in the orbital space, which facilitates subsequent discussions for
magnetic properties. Focusing on the imaginary parts of the
order parameters, we get the result 
\begin{equation}
\Phi^{(2)}=
\sum_{i}\left[
\lambda_{+}^{{\rm Im}}\eta(\Gamma'_{1},i)^{2}
+\lambda_{-}^{{\rm Im}}\sum_{l=u,v}(\eta_{l}(\Gamma'_{12},i))^{2}
\right],\label{phi2-zero}
\end{equation}
where
\begin{eqnarray*}
&&\eta(\Gamma'_{1},i)=\frac{1}{\sqrt{3}}
\sum_{a}{\rm Im}\eta_{i}(\bm{Q}_{a}),\\
&&
\eta_{u}(\Gamma'_{12},i)=\frac{1}{\sqrt{6}}\ 
{\rm Im}(2\eta_{i}(\bm{Q}_{z})
-\eta_{i}(\bm{Q}_{x})-\eta_{i}(\bm{Q}_{y})),\\
&&
\eta_{v}(\Gamma'_{12},i)=\frac{1}{\sqrt{2}}\ {\rm Im}(
\eta_{i}(\bm{Q}_{x})-\eta_{i}(\bm{Q}_{y})),
\end{eqnarray*}
transform according to the $\Gamma'_{1}$ and $\Gamma'_{12}$
representations of O$_{{\rm h}}$, respectively, under the 
rotation of the orbital space.
As $\lambda_{+}^{{\rm Im}}>\lambda_{-}^{{\rm Im}}$, 
the condensation of excitons in 
$\Gamma'_{12}$ occur, and we shall focus only on $\Gamma'_{12}$
There are two types of degeneracies in (\ref{phi2-zero}). 
One is in the summation over $i$, which reflects the SU(2) symmetry 
of the spin space. It will never be lifted in the absence of the 
spin-orbit coupling. The other is in the summation over $l$, i.e.
in the direction in the $\eta_{u}$-$\eta_{v}$ plane. This is 
lifted in the sixth order in $\eta_{l}$. To see this, let us write down 
higher order terms. They are written conveniently in terms of 
a complex order parameter $w
=\eta_{u}+{\rm i}\eta_{v}$ as
\begin{eqnarray*}
 &&\Phi=
K_{1}\sum_{i}
|w(\Gamma'_{12},i)|^{2}
+K_{2}\left(
\sum_{i}
|w(\Gamma'_{12},i)|^{2}\right)^{2}
\nonumber \\
&&\makebox[.5em]{}
+K_{3}
\left(
\sum_{i}
|w(\Gamma'_{12},i)|^{2}\right)^{3}
+K_{4}{\rm Re}\left(\sum_{i}[w(\Gamma'_{12},i)]^{2}\right)^{3}
,
\end{eqnarray*}
where $K_{j}$ are constants. This is minimized when 
\begin{enumerate}
 \item ${\rm arg}w=\frac{\pi}{3}n$\ \  if  \ $K_{4}<0$
 \item ${\rm arg}w=\frac{\pi}{6}(2n+1)$\ \  if \ $K_{4}>0$,
\end{enumerate}
where $n$ is an integer.
The order parameters are
\begin{eqnarray}
\text{(I)}&&({\rm Im}\eta_{i}(\bm{Q}_{x}),\ 
{\rm Im}\eta_{i}(\bm{Q}_{y}),\ 
{\rm Im}\eta_{i}(\bm{Q}_{z})
)=c(1,1,-2),\label{Ieta}\\
\text{(II)}&&({\rm Im}\eta_{i}(\bm{Q}_{x}),\ 
{\rm Im}\eta_{i}(\bm{Q}_{y}),\ 
{\rm Im}\eta_{i}(\bm{Q}_{z})
)=c(1,-1,0),\label{IIeta}
\end{eqnarray}
where we write down only one among three equivalent directions for each
case.
Both (I) and (II)  have tetragonal distortion.
The magnetic point group for each case is (I) $4'/{\rm m}'{\rm m}'{\rm
m}
={\rm D}_{4}\times \{E,\ RI\}$ 
(II) 4/${\rm m}'{\rm m}'{\rm m}'
={\rm D}_{2{\rm d}}\times \{E,\ RI\}
$.
The ME property tensors are easily written down \cite{birss};
\begin{equation}
 \mbox{(I)}
\left(
\begin{array}{ccc}
  G_{1}&0&0\\
  0& G_{1}&0\\
  0&0&G_{3}
\end{array}
\right),\ \ 
 \mbox{(II)}
\left(
\begin{array}{ccc}
  G_{1}&0&0\\
  0& -G_{1}&0\\
  0&0&0
\end{array}
\right),\ \ 
\end{equation}
in the cubic coordinates. Note that the ME 
effect in the absence of the spin-orbit coupling cannot originate
from spins. This shows that the orbital moment exists in this system.

It is helpful to compare the states (C-1)-(D-2) with (I)(II). 
The forms of the order parameters are given in Eqs.(\ref{C1eta})
(\ref{D1eta}) for (C-1) and (D-1), and in Eqs.(\ref{Ieta}) (\ref{IIeta}) for 
(I)(II).
Naively one may expect that (C-1)-(D-2) are included  in (I) or (II), because
(C-1)-(D-2) assumes nonzero spin-orbit coupling, whereas (I)(II)
assumes an absence of the spin-orbit coupling.
It is, however, not the case, as seen from the different ME tensors in 
the two cases. The reason is the following. In the 
presence of the spin-orbit coupling, the degeneracy in the GL free energy 
$\Phi$ is lifted in the quartic order, resulting in (C-1)-(D-2). 
In contrast, without the spin-orbit coupling, the degeneracy is lifted 
in the sixth order, leading to (I)(II). Because of this difference
in the lifting of degeneracy, the realized states are different 
between the cases with and without the spin-orbit coupling.


\begin{thebibliography}{99}
\bibitem{young}
D.~P.~Young, D.~Hall, M.~E.~Torelli, Z.~Fisk, J.~L.~Sarrao, 
J.~D.~Thompson, H.~R.~Ott, S.~B.~Oseroff, R.~G.~Goodrich, 
and R.~Zysler, Nature {\bf 397} 412 (1999).
\bibitem{ceperley}
D.~Ceperley, Nature {\bf 397} 386 (1999).
\bibitem{zra}
M.~E.~Zhitomirsky, T.~M.~Rice and V.~I.~Anisimov, Nature {\bf 402} 251 
(1999).
\bibitem{bv}
L.~Balents and C.~M.~Varma, Phys. Rev. Lett. {\bf 84} 1264 (2000).
\bibitem{vkr}
B.~A.~Volkov, Yu.~V.~Kopaev and A.~I.~Rusinov, Sov. Phys. JETP 
{\bf 41} 952 (1976).
\bibitem{hy}
A.~Hasegawa and A.~Yanase, J. Phys. {\bf C12} 
5431 (1979).
\bibitem{mcpm}
S.~Massida, A.~Continenza, T.~M.~de~Pascale and R.~Monnier, 
Z. Phys. {\bf B102} 83 (1997).
\bibitem{hall}
D.~Hall, D.~P.~Young, Z.~Fisk, T.~P.~Murphy, E.~C.~Palm, A.~Teklu,
and R. G. Goodrich,
Phys. Rev. {\bf B64} 233105 (2001).
\bibitem{hr}
B.~I.~Halperin and T.~M.~Rice, in {\it Solid State Physics} {\bf 21} 115
(eds. F.~Seitz, D.~Turnball, and H.~Ehrenfest, Academic Press, New York 1968).
\bibitem{tromp}
H.~J.~Tromp, P.~van~Gelderen, P.~J.~Kelly, G.~Brocks, and 
P.~A.~Bobbert, Phys. Rev. Lett. {\bf 87}, 016401 (2001):
in GW, $G$ and $W$ represents the Green's function and the screened Coulomb 
potential, respectively. The self-energy is given by the product of these
two.
\bibitem{ARPES}
J.~D.~Denlinger, J.~A.~Clack, J.~W.~Allen, G.~H.~Gweon, D.~M.~Poirier, 
C.~G.~Olson, J.~L.~Sarrao, A.~D.~Bianchi, and Z.~Fisk, 
preprint (cond-mat/0107429), to appear in Phys. Rev. Lett.
\bibitem{Xray}
J.~D.~Denlinger, G.~H.~Gweon, J.~W.~Allen, A.~D.~Bianchi, and 
Z.~Fisk, preprint (cond-mat/0107426).
\bibitem{kino}
H.~Kino, F.~Aryasetiawan, T.~Miyake,
and K.~Terakura, preprint.
\bibitem{udagawa}
M.~Udagawa, S.~Nagai, N.~Ogita, F.~Iga, 
R.~Kaji, K.~Sumida, J.~Akimitsu, and S.~Kunii, K.~Suzuki, H.~Onodera, and 
Y.~Yamaguchi,
J. Phys. Soc. Jpn. Suppl. {\bf 71} 314 (2002).
\bibitem{ohishi}
K.~Ohishi, T.~Yokoo, K.~Kakuta, H.~Takigawa, A.~Tagaya, K.~Takenawa, R.~Kaji,
J.~Akimitsu, W.~Higemoto, and R.~Kadono,
Newsletter of Scientific Research on Priority 
Areas (B) {\it Orbital Orderings and Fluctuations}, Vol.~1, No.~2 (2000)    
16.
\bibitem{bg}
V.~Barzykin and L.~P.~Gor'kov, Phys. Rev. Lett. {\bf 84} 2207 (2000).
\bibitem{b}
L.~Balents, Phys. Rev. {\bf B62} 2346 (2000)
\bibitem{vb}
M.~Y.~Veillette and L.~Balents, Phys. Rev. {\bf B65} 014428 (2001)
\bibitem{bbm}
E.~Bascones, A.~A.~Burkov, and A.~H.~MacDonald, Phys. Rev. Lett. {\bf 89}
086401 (2002).
\bibitem{morikawa}
T.~Morikawa, T.~Nishioka and N.~K.~Sato, J. Phys. Soc. Jpn. {\bf 70}
341 (2001).
\bibitem{ott}
H.~R.~Ott, J.~L.~Gavilano, B.~Ambrosini, P.~Vonlanthen, E.~Felder, 
L.~Degiorgi, D.~P.~Young, Z.~Fisk, and R.~Zysler, 
Physica {\bf 281B-282B} 423 (2000).
\bibitem{terashima} T.~Terashima, (private communication). 
\bibitem{kunii}
S.~Kunii, J. Phys. Soc. Jpn. {\bf 69} 3789 (2000).
\bibitem{msnm}
S.~Murakami, R.~Shindou, N.~Nagaosa, and A.~S.~Mishchenko,
Phys. Rev. Lett. {\bf 88} 126404 (2002).
\bibitem{hrremark}
This matrix $\bm{\eta}$ corresponds to the transpose of the 
matrix $\bm{M}$ in Ref.~\onlinecite{hr}. 
\bibitem{vg}
G.~E.~Volovik and L.~P.~Gor{'}kov, Sov. Phys. JETP {\bf 61} 843 (1985).
\bibitem{su}
M.~Sigrist and K.~Ueda, Rev. Mod. Phys. {\bf 63} 239 (1991).
\bibitem{gdb6}
H.~Nozaki, T.~Tanaka, and Y.~Ishizawa, J. Phys. C {\bf 13} 2751 (1980)
\bibitem{ceb6}
J.~M.~Effantin
J.~Rossat-Mignod, P.~Burlet, H.~Bartholin, S.~Kunii and T.~Kasuya, 
J. Magn. Magn. Mater. {\bf 47\&48} 145 (1985).
\bibitem{eub6}
S.~S{\"u}llow, I.~Prasad, M.~C.~Aronson, J.~L.~Sarrao,
Z.~Fisk, D.~Hristova, A.~H.~Lacerda, M.~F.~Hundley,
A.~Vigliante, and D.~Gibbs, Phys. Rev. {\bf B57} 5860 (1998).
\bibitem{cubic}
W.~Opechowski and R.~Guccione in {\it Magnetism} (eds. G.~T.~Rado and 
H.~Suhl, Academic Press, New York, 1965) Vol.~IIa, p.~105.
\bibitem{birss}
R.~R.~Birss, {\it Symmetry and Magnetism} pp.~136-145 
(ed. E.~P.~Wohlfarth, North
Holland, Amsterdam, 1964).
\bibitem{hr-comment}
The pure imaginary order parameters in this letter correspond to
Class II and III in Ref.~\onlinecite{hr}, which break time-reversal symmetry. 
Likewise the real parts correspond to 
Class I and IV in 
Ref.~\onlinecite{hr}, with time-reversal symmetry being preserved.
Thus our observation that the order parameters become pure imaginary 
coincides with the result in Ref.~\onlinecite{hr} that the mixture of Classes 
II and III is the most 
favorable state in the presence of spin-orbit coupling, provided the 
electron-phonon coupling is not so strong.
\bibitem{udagawa2}
M.~Udagawa, (private communication).
\bibitem{dzy1}
I.~E.~Dzyaloshinskii, Sov. Phys. JETP {\bf 5} 1259 (1957).
\bibitem{PMnote}
If intervalley excitons 
condense, i.e. the assumption (ii) is violated, there will be 
a possibility of the PM effect.
\bibitem{dzy2}
I.~E.~Dzyaloshinskii, Sov. Phys. JETP {\bf 10} 628 (1960).
\bibitem{odell}
{\it The Electrodynamics of Magneto-electric Media} (T.~H.~O'Dell, 
North-Holland, Amsterdam, 1970).
\bibitem{saitoh}
M.~Saitoh, H.~Takigawa, H.~Ichikawa, T.~Yokoo, J.~Akimitsu, 
M.~Nishi, K.~Kakurai, M.~Takata, N.~Okada, M.~Sakata, and S.~Kunii,
J. Phys. Soc. Jpn. Suppl. {\bf 71}, 106 (2002).
\bibitem{fukuyama}
H.~Fukuyama and R.~Kubo, J. Phys. Soc. Jpn. {\bf 28}, 570 (1970).
\bibitem{ffsp}
M.~Fiebig, D.~Fr{\"o}hlich, G. Sluyterman v. L., and R.~V.~Pisarev,
Appl. Phys. Lett. {\bf 66}, 2906 (1995).
\bibitem{bst}
W.~F.~Brown,  S.~Shtrikman and D.~Treves, J. Appl. Phys. {\bf 34}, 
1233 (1963).
\bibitem{fuchs}
R.~Fuchs, Phil. Mag. {\bf 11}, 647 (1965).
\bibitem{gr92}
E.~B.~Graham and R.~E.~Raab, Phil. Mag. {\bf B66}, 269 (1992).
\bibitem{gr99}
E.~B.~Graham and R.~E.~Raab, Phys. Rev. {\bf B59}, 7058 (1999).
\bibitem{rikken}
G.~L.~J.~A.~Rikken and R.~Raupach, Nature {\bf 390}, 493 (1997).
\bibitem{x}
J.~Goulon, A.~Rogalev, C.~Goulon-Ginet, G.~Benayoun, L.~Paolasini,
C.~Brouder, C.~Malgrange, and P.~A.~Metcalf, 
Phys. Rev. Lett. {\bf 85}, 4385
(2000).
\bibitem{dp}
I.~Dzyaloshinskii and E.~V.~Papamichail, Phys. Rev. Lett. {\bf 75},
3004 (1995).
\bibitem{ffkp}
M.~Fiebig, D.~Fr{\"o}hlich, B.~B.~Krichevtsov, and R.~V.~Pisarev,
Phys. Rev. Lett. {\bf 73}, 2127 (1994).
\bibitem{YMO98}
D.~Fr{\"o}hlich, Th.~Kiefer, St.~Leute, Th.~Lottermoser,
Appl. Phys. {\bf B68}, 465 (1999).
\bibitem{YMO99}
D.~Fr{\"o}hlich, St.~Leute, V.~V.~Pavlov, R.~V.~Pisarev,
Phys. Rev. Lett. {\bf 81}, 3239 (1998).
\bibitem{spin-flop}
M.~Fiebig, D.~Fr{\"o}hlich, H.~-J.~Thiele,
Phys. Rev. {\bf B54}, R12681 (1996).
\bibitem{phi}
This tilting angle should not be so large as to avoid an 
affect of birefringence. In YMnO$_3$ it is discussed in 
Ref.~\onlinecite{YMO99}
that 
this $\phi$ should be less than 20 degrees.
\bibitem{YMO00}
M.~Fiebig,
D. Fr{\"o}hlich, K.~Kohn, St.~Leute, Th.~Lottermoser, V.~V.~Pavlov,
and R.~V.~Pisarev,
Phys. Rev. Lett. {\bf 84}, 5620 (2000).
\bibitem{gavilano}
J.~L.~Gavilano, Sh.~Mushkolaj, D.~Rau, H.~R.~Ott, A.~Bianchi, 
D.~P.~Young, and Z.~Fisk, Phys. Rev. {\bf B63} 140410 (2001).
\bibitem{gianno}
K.~Gianno', A.~V.~Sologubenko, H.~R.~Ott, A.~D.~Bianchi
Z.~Fisk, preprint (cond-mat/0104511).
\bibitem{sk}
D.~Sherrington and W.~Kohn, Rev. Mod. Phys. {\bf 40} 767 (1968).
\bibitem{z}
J.~Zittartz, Phys. Rev. {\bf 164} 575 (1967).
\bibitem{md}
R.~Monnier and B.~Delley, Phys. Rev. Lett.{\bf 87} 157204 (2001).
\bibitem{akimitsu}
J.~Akimitsu, K.~Takenawa, K.~Suzuki, H.~Harima, and 
Y.~Kuramoto, Science {\bf 293} 1125 (2001).
\bibitem{makarova}
T.~L.~Makarova, B.~Sundqvist, R.~H{\"o}hne, P.~Esquinazi, 
Y.~Kopelevich, P.~Scharff, V.~A.~Davydov, L.~S.~Kashevarova,
and A.~V.~Rakhmanina, Nature {\bf 413} 716 (2001).
\bibitem{murakami}
S.~Murakami, (unpublished).
\bibitem{shen}
Y.~R.~Shen, {\it The Principles of Nonlinear Optics} (Wiley, New York, 1984).
\end{thebibliography}
\end{document}